\documentclass{aa}
\usepackage[dvips]{graphicx}
\newcommand{\gtrsim}{\mathrel{\hbox{\rlap{\hbox{\lower4pt\hbox{$\sim$}}}\hbox{$>$}}}}
\newcommand{\lesssim}{\mathrel{\hbox{\rlap{\hbox{\lower4pt\hbox{$\sim$}}}\hbox{$<$}}}}

\def\MCN{\mbox{CH$_3$CN}}
\def\MCNI{$^{13}$CH$_3$CN}
\def\MCNII{CH${_3}^{13}$CN}

\def\HII{H{\sc ii} }

\def\HC{HC~H{\sc ii}}

\def\kms{\mbox{km~s$^{-1}$}}

\def\Vlsr{$V_{\rm LSR}$}

\def\mjy{~mJy~beam$^{-1}$}

\def\jdo{\mbox{12--11}}
\def\jcc{\mbox{5--4}}
\def\jdu{\mbox{2--1}}
\def\juz{\mbox{1--0}}

\begin{document}
\title{Molecular outflows and hot molecular cores in G24.78+0.08 at sub-arcsecond
angular resolution} 
\author{M.\ T.\ Beltr\'an\inst{1} \and R.\ Cesaroni \inst{1} \and Q.\ Zhang \inst{2} \and
R.\ Galv\'an-Madrid \inst{2,3,4} \and H.\ Beuther \inst{5} \and C.\ Fallscheer
\inst{5,6} \and
R.\ Neri \inst{7}
\and C.\ Codella \inst{1}}
\institute{
INAF, Osservatorio Astrofisico di Arcetri, Largo E.\ Fermi 5,
50125 Firenze, Italy
\and
Harvard-Smithsonian Center for Astrophysics, 60 Garden Street, Cambridge MA 02138, USA
\and
Centro de Radioastronom{\'\i}a y Astrof{\'\i}sica, Universidad Nacional Aut\'onoma de
M\'exico, Morelia 58090, Mexico 
\and
Academia Sinica Institute of Astronomy and Astrophysics, Taipei 106, Taiwan
\and
Max Planck Institute for Astronomy, K\"onigstuhl 17, 69117 Heidelberg, Germany
\and
Department of Physics and Astronomy, University of Victoria, 3800 Finnerty Road, Victoria, BC V8P 5C2, Canada
\and
IRAM, 300 Rue de la Piscine, F-38406 Saint Martin d'H\`eres, France} 


\offprints{M.\ T.\ Beltr\'an, \email{mbeltran@arcetri.astro.it}}
\date{Received date; accepted date}

\titlerunning{G24.78+0.08}
\authorrunning{Beltr\'an et al.}

\abstract
{This study is part of a large project to study the physics of accretion and molecular
outflows towards a selected sample of high-mass star-forming regions that show evidence of
infall and rotation from previous studies.}
{We wish to make a thorough study at high-angular resolution of the structure and
kinematics of the HMCs and corresponding molecular outflows in the high-mass star-forming 
region G24.78+0.08.}
{We carried out SMA and IRAM PdBI observations at 1.3 and 1.4~mm, respectively, of dust
and of typical high-density and molecular outflow tracers with resolutions of $<1"$.
Complementary IRAM 30-m $^{12}$CO and $^{13}$CO observations were carried out to recover
the short spacing information of the molecular outflows.}
{The millimeter continuum emission towards cores G24 A1 and A2 has been resolved into 3
and 2 cores, respectively, and  named A1, A1b, A1c, A2, and A2b. All these cores are
aligned in a southeast-northwest direction coincident with that of the molecular outflows
detected in the region, which suggests a preferential direction for star formation in this
region. The masses of the cores range from 7 to 22~$M_\odot$, and the rotational
temperatures from 128 to 180~K. The high-density tracers have revealed the existence of 2
velocity components towards A1, one of them peaks  close to the position of the
millimeter continuum peak and of the \HC\ region, and is associated  with the velocity
gradient seen in \MCN\ towards this core, while the other one peaks southwest of core
A1 and is not associated with any millimeter continuum emission peak. The position-velocity plots along outflow A and the $^{13}$CO~(\jdu) averaged blueshifted
and redshifted emission indicate that this outflow is driven by core A2. Core A1
apparently is not driving any outflow. The knotty appearance
of the highly collimated outflow C and the $^{12}$CO position-velocity plot suggest an
episodic outflow, where the knots are made of swept-up ambient gas.}
{}

\maketitle

\section{Introduction}

Molecular outflows, infall and rotation are ubiquitous phenomena during the
earliest stages of the formation of stars of all masses and luminosities. During
the last few years, much effort has been devoted to studying and describing the
physical properties of embedded massive protostars and their molecular outflows 
(e.g., Shepherd \& Churchwell 1996a,b; Zhang et al.~\cite{zhang01}; Beuther et
al.~\cite{beuther02}; L\'opez-Sepulcre~\cite{lopez09}). The presence of rotating
structures around massive young stars has also been asserted thanks to the
detection of velocity gradients perpendicular to the axis of the molecular
outflows in the cores (see Cesaroni~\cite{cesa07} and reference therein). These
studies have revealed two different kinds of structures surrounding B- and
O-type stars. On the one hand, B-type stars are surrounded by centrifugally
supported disks, as seen in the prototypical B-type protostar IRAS~20126+4104
(Cesaroni et al.~\cite{cesa05}; Keto \& Zhang~\cite{keto10}). On the other hand,
despite the fact that theory predicts the existence of both  large rotating
structures (e.g.\ Peters et al.~\cite{peters10}) and accretion disks (e.g.\
Krumholz et al.~\cite{krumholz09}) around massive O-type stars, such disks have
been elusive to date. One finds, instead, huge ($\sim$0.1 pc), massive (a few
100~$M_\odot$) rotating toroids, which are non Keplerian and very likely
non-equilibrium structures (Beltr\'an et al.~\cite{beltran11}). Finally,  infall
onto young O-type stars has been measured through high-angular resolution
observations of lines showing redshifted self-absorption or absorption against a
background bright continuum source (Keto et al.~\cite{keto87}; Zhang et
al.~\cite{zhang98}; Beltr\'an et al.~\cite{beltran06}), or by measuring the
velocity field of the ionized gas surrounding a newly formed massive star (see
Keto~\cite{keto02}). 

\begin{table*}
\caption[] {Parameters of the maps}
\label{par_obs}
\begin{tabular}{lcccccc}
\hline
&&&\multicolumn{2}{c}{Synthesized beam$^{\rm a}$}  
\\
\cline{4-5}
&&\multicolumn{1}{c}{Frequency} &
\multicolumn{1}{c}{$HPBW$} &
\multicolumn{1}{c}{P.A.}&
\multicolumn{1}{c}{Spectral resolution}&
\multicolumn{1}{c}{rms noise$^{\rm b}$}
\\
\multicolumn{1}{c}{Observation} &
\multicolumn{1}{c}{Telescope$^{\rm a}$}&
\multicolumn{1}{c}{(GHz)} &
\multicolumn{1}{c}{(arcsec)} &
\multicolumn{1}{c}{(deg)}  &
\multicolumn{1}{c}{(\kms)} &
\multicolumn{1}{c}{(mJy~beam$^{-1}$)} \\
\hline
continuum                                &SMA &225.31 &$0.95\times0.75^{\rm c}$  &61$^{\rm c}$ &-- &3.0$^{\rm c}$  \\
continuum                                &PdBI &214.30 &$0.89\times0.29$ &10 &-- &3.2 \\ 
$^{13}$CH$_3$CN (12$_K$--11$_K$)         &PdBI &214.374$^{\rm d}$ &$0.89\times0.29^{\rm e}$&10 &0.25 &15\\
C$^{18}$O (\jdu)                         &SMA &219.560 &$0.83\times0.64^{\rm e}$ &55  &0.6 &35 \\
HNCO (10$_{0,10}$--9$_{0,9}$)            &SMA &219.798 &$0.72\times0.51$ &46 &0.6 &45 \\
H$_2$$^{13}$CO (3$_{1,2}$--2$_{1,1}$)    &SMA &219.909 &$0.72\times0.51$ &46 &0.6 &45 \\
SO (6$_5$--5$_4$)                        &SMA &219.949 &$0.83\times0.64^{\rm e}$ &55 &0.6 &40 \\
CH$_3$OH (8$_{0,8}$--7$_{1,6}$)          &SMA &220.079 &$0.66\times0.48$ &48 &0.6 &45 \\
HCOOCH$_3$ (17$_{2,15}$--16$_{4,2}$)     &SMA &220.167 &$0.71\times0.50$ &45 &0.6 &45 \\
CH$_2$CO (1$_{1,11}$--10$_{1,10}$)       &SMA &220.178 &$0.71\times0.50$ &45 &0.6 &45 \\
HCOOCH$_3$ (17$_{4,13}$--16$_{4,12}$)    &SMA &220.190 &$0.71\times0.50$ &45 &0.6 &45 \\
$^{13}$CO (\jdu)                         &SMA &220.399 &$0.83\times0.64^{\rm e}$ &55 &1.0 &35 \\
$^{13}$CO (\jdu)                         &SMA+IRAM~30-m &220.399 &$3.57\times2.02$ &71 &1.0 &116 \\
\MCNII\ (12$_K$--11$_K$)                 &SMA &220.638$^{\rm d}$ &$0.67\times0.47$ &47 &0.6 &45\\
C$_2$H$_5$CN (25$_{2,24}$--24$_{2,23}$)  &SMA &220.661 &$0.67\times0.48$ &47 &0.6 &45 \\
\MCN\ (12$_K$--11$_K$)                   &SMA &220.747$^{\rm d}$ &$0.71\times0.50$ &46 &0.6 &40\\
\MCN\ $v_8=1$ ( $K,l$=6,$+1$)            &SMA &221.312  &$0.66\times0.47$ &46 &0.6 &40 \\
\MCN\ $v_8=1$ ($K,l$=3,$-1$)             &SMA &221.338 &$0.66\times0.47$ &46 &0.6 &40 \\
CH$_3$OH (8$_{1,8}$--7$_{0,7}$)E         &SMA &229.759 &$0.65\times0.47$ &48 &0.6 &50 \\
CH$_3$OH (19$_{5,14}$--20$_{4,17}$)A$-$  &SMA &229.939 &$0.65\times0.47$ &47 &0.6 &45 \\
CH$_3$OH (3$_{2,2}$--4$_{1,4}$)E         &SMA &230.027 &$0.65\times0.47$ &48 &0.6 &45 \\
$^{12}$CO (\jdu)                         &SMA &230.538 &$0.86\times0.68^{\rm e}$ &54 &1.5 &40 \\
$^{12}$CO (\jdu)                         &SMA+IRAM~30-m &230.538 &$3.37\times1.96$ &70 &1.0 &116 \\
OCS (19--18)                             &SMA &231.061 &$0.69\times0.50$ &46 &0.6 &50 \\
$^{13}$CS (5--4)                         &SMA &231.221 &$0.70\times0.49$ &46 &0.6 &50 \\       
\hline
\end{tabular}

$^a$ The synthesized CLEANed beams for maps with the ROBUST parameter of
Briggs~(\cite{briggs95}) set equal to 0.\\
$^b$ For the molecular line observations the 1$\sigma$ noise is per channel.\\
$^c$ The synthesized beam for the SMA VEX configuration map  
is $0\farcs55\times0\farcs40$ at
P.A.=52$\degr$, and the rms noise is 3.2~mJy~beam$^{-1}$. \\
$^d$ The frequency is that of the $K$=0 component. \\
$^e$ The synthesized CLEANed beams for maps using natural weighting. 
\end{table*}

G24.78+0.08 (hereafter G24) is one of the most studied massive star-forming regions. The
simultaneous presence of the three elements fundamental for star formation (outflow,
infall, and rotation) has been reported in one of the massive objects embedded in this
clump. This region, which is located at a distance of 7.7~kpc, has been studied in great
detail by our group (Codella et al.~\cite{codella97}; Furuya et al.~\cite{furuya02};
Cesaroni et al.~\cite{cesa03}; Beltr\'an et al.~\cite{beltran04}, Beltr\'an et
al.~\cite{beltran05}, Beltr\'an et al.~\cite{beltran06}, Beltr\'an et
al.~\cite{beltran07}; Moscadelli et al.~\cite{mosca07}; Vig et al.~\cite{vig08}). G24
contains a cluster of Young Stellar Objects (YSOs), three of which have rotating toroids
associated with them (Beltr\'an et al.~\cite{beltran04}). These rotating toroids seem to
be associated with two compact molecular outflows, first mapped in CO by  Furuya et
al.~(\cite{furuya02}). In one of these rotating cores, G24~A1, the presence of an
early-type star is witnessed by an embedded hypercompact (HC) \HII region, first detected
at 1.3~cm by Codella et al.~(\cite{codella97}) and later studied at very high-angular
resolution at 1.3 and 0.7~cm by Beltr\'an et al.~(\cite{beltran07}). The properties of the
ionized gas have also been studied through continuum emission and recombination lines by
Galv\'an-Madrid et al.~(\cite{galvan08}) and Longmore et al.~(\cite{longmore09}). This
\HC\ region, which has a diameter of $\la$$0\farcs2$ ($\la$1,500~AU), is located at the 
geometrical center of the toroid. The free-free continuum spectrum resembles that of a
classical (Str\"omgren) \HII region around a zero-age main sequence star of spectral type
O9.5, corresponding to a mass of $\sim 20~M_\odot$. Beltr\'an et al.~(\cite{beltran06})
mapped the core G24~A1 in NH$_3$ and detected redshifted absorption, indicating that the
toroid is undergoing infall towards the \HC\ region.

We have carried out new 1.3~mm interferometric observations of this region at
sub-arcsecond resolution with the Submillimeter Array (SMA)\footnote{The Submillimeter Array is a joint project between the Smithsonian Astrophysical Observatory and the Academia Sinica
Institute of Astronomy and Astrophysics, and is funded by the Smithsonian Institution and
the Academia Sinica.} (Ho et al.~\cite{ho04}) of several outflow and high-density
tracers. The goals of this study were to better characterize the structure and kinematics
of the Hot Molecular Cores (HMCs) and molecular outflows in this region, and in
particular, to identify the powering source of the outflow observed towards G24~A1. In
order to be sensitive to extended structures filtered out by the interferometer, we have
also mapped the region over $2'$ with the IRAM 30-m telescope\footnote{IRAM is supported by INSU/CNRS
(France), MPG (Germany) and IGN (Spain).}. In addition, we used the
enhanced capabilities of the IRAM Plateau de Bure (PdBI) to observe this region in
\MCNI~(\jdo) at very high-angular resolution to investigate whether the massive toroids
were hiding true (Keplerian) circumstellar disks in their interiors. Unfortunately, the
angular resolution of the observations has not been enough to detect circumstellar disks
embedded in the massive toroids, but it has allowed us to study the velocity structure of
the cores with unprecedented resolution. The observational details are given in Sect.~2,
while the results are illustrated in Sect.~3 and analyzed and discussed in Sect.~4.
Finally, the conclusions are drawn in Sect.~5.

\begin{figure*}
\centerline{\includegraphics[angle=-90,width=16cm]{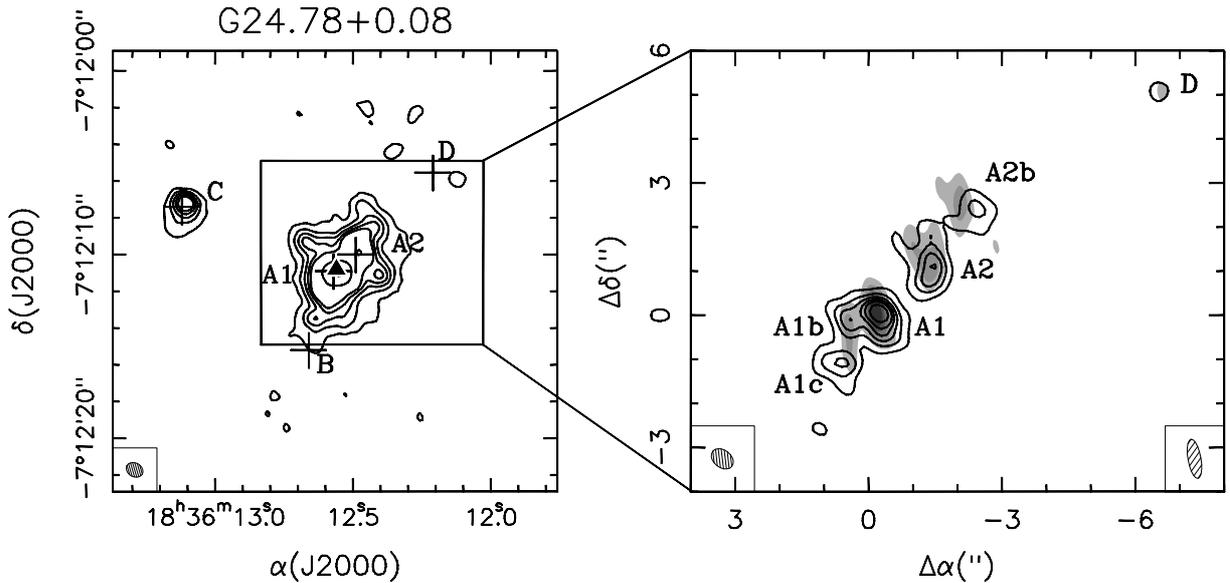}}
\caption{({\it Left panel}) SMA map of the 1.3~mm continuum emission obtained combining both
compact and very extended configurations.  Contour levels are 3, 6, 9, 12, 15,
35, and
60 times  $\sigma$, where 1$\sigma$ is 3\mjy.  The crosses mark the positions of the
sources identified by Furuya et al.~(\cite{furuya02}) and Beltr\'an et al.~(\cite{beltran04}).
The  triangle marks the position of the \HC\ region in G24 A1 (Beltr\'an et
al.~\cite{beltran07}). The SMA synthesized beam is shown in the lower left corner. 
({\it Right panel}) Very extended configuration only SMA map at 1.3~mm ({\it contours}) overlaid on
the PdBI map at 1.4~mm ({\it grayscale}) towards G24 A1, A2 and D. Contour and grayscale levels are 3, 6, 9, 12, 15, 
and 20 times $\sigma$, where 1$\sigma$ is
3.2\mjy. The SMA synthesized beam is shown in the lower left corner and the PdBI one in the lower right corner. }
\label{g_cont}
\end{figure*}

\section{Observations}

\subsection{Submillimeter Array}

G24 was observed with the SMA using two different array configurations.
Compact-array observations were taken on July 17, 2007, and covered baselines with lengths
between 7 and 100~k$\lambda$ (sampling spatial structures in the range of $29\farcs5$ to
$2\farcs1$). Very Extended (VEX) configuration data were made on August 2, 2008, with
baseline lengths from 23 to 391~k$\lambda$ ($9\farcs0$ to $0\farcs5$). The SMA had two
spectral sidebands - both 2~GHz wide - separated by 10~GHz. For both observations, the two
sidebands covered the frequency ranges of 219.3--221.3 and 229.3--231.3~GHz with a
uniform spectral resolution of $\sim$0.5~\kms.

The phase reference center of the observations was set to the position
$\alpha$(J2000)=18$^{\rm h}$ 36$^{\rm m}$ $12\fs565$, $\delta$(J2000)=$-$07$''$ 12$'$
$10\farcs90$. Absolute flux calibration was derived from observations of Callisto. The
bandpass of the receiver was calibrated by observations of the quasar  3C454.3. Amplitude
and phase calibrations were achieved by monitoring 1733--130. The flux densities estimated
for 1733$-$130 are 1.71 and 2.29~Jy for the compact and the VEX configuration,
respectively. We estimate the flux-scale uncertainty to be better than 15\%. The
visibilities were calibrated with the IDL superset MIR\footnote{The MIR cookbook by
Charlie Qi can be found at http:// cfa-www.harvard.edu/$\sim$cqi/mircook.html}. Further imaging
and analysis was done with MIRIAD (Sault et al.~\cite{sault95}) and  GILDAS\footnote{The
GILDAS package is available at http://www.iram.fr/IRAMFR/GILDAS}. The continuum was
constructed in the {\it(u,v)}-domain from the line-free channels.  Continuum and channel
maps were created combining the  data of  both compact and VEX configurations with the
ROBUST parameter of Briggs~(\cite{briggs95}) set equal to 0, except for the $^{12}$CO,
$^{13}$CO, C$^{18}$O, and SO channel maps that were created using natural weighting.  
Details of the synthesized CLEANed beam, spectral resolution, and rms noise of the maps 
obtained in a number of lines are given in Table~\ref{par_obs}.


\begin{table*}
\caption[] {Parameters of the cores}
\label{table_cont}
\begin{tabular}{lcccccccc}
\hline
&\multicolumn{2}{c}{Position peak$^a$}
&&&\multicolumn{2}{c}{Source Diameter$^{a, b}$}
\\
 \cline{2-3} 
 \cline{6-7} 
&\multicolumn{1}{c}{$\alpha({\rm J2000})$} &
\multicolumn{1}{c}{$\delta({\rm J2000})$} &
\multicolumn{1}{c}{$I^{\rm peak}_\nu$} &
\multicolumn{1}{c}{$S_\nu$} &
\multicolumn{1}{c}{$\theta_s$} &
\multicolumn{1}{c}{$\theta_s$} &
\multicolumn{1}{c}{$T_{\rm rot}$\,$^c$} &
\multicolumn{1}{c}{$M_{\rm gas}$\,$^{d}$} 
\\
\multicolumn{1}{c}{Core} &
\multicolumn{1}{c}{h m s}&
\multicolumn{1}{c}{$\degr$ $\arcmin$ $\arcsec$} &
\multicolumn{1}{c}{(mJy/beam)} & 
\multicolumn{1}{c}{(mJy)} &
\multicolumn{1}{c}{(arcsec)} &
\multicolumn{1}{c}{(AU)} &
\multicolumn{1}{c}{(K)} &
\multicolumn{1}{c}{($M_\odot$)} 
\\
\hline
&&&~~~~VEX  \\
\hline
G24 A1   &18 36 12.552 &$-$07 12 10.80  &86.1   &187   &0.47  &3620 &148     &16$^e$ \\
G24 A1b  &18 36 12.592 &$-$07 12 11.00  &38.8   &57    &0.31  &2390 &180     &7\\
G24 A1c  &18 36 12.606 &$-$07 12 12.00  &31.2   &64    &0.65  &5000 &146     &10\\ 
G24 A2   &18 36 12.464 &$-$07 12 09.80  &38.9   &116   &0.69  &5315 &128     &22\\
G24 A2b  &18 36 12.404 &$-$07 12 08.50  &24.7   &39    &0.49  &3770 &30--60  &16--36\\
G24 C    &18 36 13.096 &$-$07 12 07.10  &41.6   &65    &0.38  &2925 &30--60  &27--60\\
G24 D    &18 36 12.128 &$-$07 12 05.80  &13.6   &14    &0.37  &2850 &30--60  &6--13 \\
\hline
&&&~~~~VEX+compact  \\
\hline
G24 A1+A1b+A1c &&&196 &865 &&&148 &125$^f$\\
G24 A2+A2b &&&107 &450 &&&128 &84\\
G24 C &&&64 &140  &&&30--60 &27--59\\
G24 D &&&16 &14.5 &&&30--60 &7--15 \\
\hline

\end{tabular}
   
  $^a$ The positions and diameters have been estimated from the very extended configuration SMA map. \\
  $^b$ Deconvolved average diameter of the 50\% contour level. \\
  $^c$ From the rotational diagram method (see Sect.~\ref{rot}). For those cores
  for which an estimate of $T_{\rm rot}$ is not possible with this method, a
  range of temperatures is given.\\
  $^d$ Mass estimated using a dust temperature equal to $T_{\rm rot}$ and a dust
  opacity of 0.8~cm$^{2}$\,g$^{-1}$ at 1.3~mm (see Sect.~\ref{rot}).  \\
  $^e$ The dust continuum emission associated with core A1 and used to estimate the
  mass is 102~mJy (see Sect.~\ref{continuum}). \\
  $^f$ The dust continuum emission associated with cores A1+A1b+A1c and used to estimate the
  mass is 780~mJy (see Sect.~\ref{continuum}). \\
\end{table*}

\subsection{IRAM Plateau de Bure Interferometer}

PdBI observations at 214.3~GHz in the B
configuration were carried out on March 9 and 11, 2007. The correlator was set up to
observe the $^{13}$CH$_3$CN~(12--11) line emission at 216~GHz with both polarizations. 
The units of the correlator were placed in such a way that a frequency range free of
lines could be used to measure the continuum flux. A unit of 80~MHz of bandwidth was set
to cover the emission of \MCNI~(\jdo) $K=$0 to 4, with a spectral resolution of 0.156~MHz
or 0.22~\kms. A second unit of 320~MHz of bandwidth was set to cover the emission of
\MCNI~(\jdo) $K=$0 to 7, with a spectral resolution of 2.5~MHz or $\sim$3.5~\kms. The two
remaining units of 320~MHz of bandwidth were placed to cover the rest of the band.


The phase center used was $\alpha$(J2000)=18$^{\rm h}$ 36$^{\rm m}$ $12\fs661$, 
$\delta$(J2000)=$-$07$''$ 12$'$ $10\farcs15$. The bandpass of the receivers was
calibrated by observations of the quasar 3C273. Amplitude and phase calibrations were
achieved by monitoring 1741$-$038, whose flux densities were determined relative to
MWC349 or 1749+096. The flux densities estimated for 1741$-$038 are 1.18 and 1.12~Jy for
March 9 and 11, respectively.  The uncertainty in the amplitude calibration is estimated
to be $\sim$20\%. The data were calibrated and analyzed with the GILDAS software package.
The continuum maps were created from the line free channels. We subtracted the continuum
from the line emission directly in the {\it(u,v)}-domain. Continuum and channel maps 
were created using natural weighting. Details of the synthesized CLEANed beam, spectral
resolution, and rms noise of maps are given in
Table~\ref{par_obs}. 

\subsection{IRAM 30-m telescope}

Supplementary short-spacing information to complement the interferometric data was
obtained with the IRAM~30-m telescope. The observations were carried out on November 11th,
2006 with the HERA receiver using the On-The-Fly mode. The receiver was tuned to
230.538~GHz for $^{12}$CO\,(\jdu) (HERA1) and 220.399~GHz for $^{13}$CO\,(\jdu) (HERA2). Both
lines were observed simultaneously and covered with the VESPA
autocorrelator with 0.1~\kms\ spectral resolution, corresponding to a channel spacing of
$\sim$0.07~MHz. The maps sizes were $2\arcmin\times2\arcmin$ and were centered at 
$\alpha$(J2000)=18$^{\rm h}$36$^{\rm m}$12\fs6 and
$\delta$(J2000)=--07\degr12\arcmin10\farcs9. The sampling interval was 2$''$ and the
region was alternatively scanned in the north-south direction and the east-west direction
in order to reduce effects caused by the scanning process. 

The spectra were reduced using CLASS of the GILDAS software package and channel maps with
1~\kms\ resolution were created. The $^{12}$CO data have a beam size of $10\farcs4$ and
the $^{13}$CO of $10\farcs9$. The IRAM 30-m data sets were combined with the SMA data sets
using the UVMODEL task of the MIRIAD package. The synthesized beam of the combined
SMA+IRAM~30-m data is $3\farcs37\times1\farcs96$ at P.A.=70$\degr$ for $^{12}$CO and
$3\farcs57\times2\farcs02$ at P.A.=71$\degr$ for $^{13}$CO

\section{Results}

\subsection{Continuum emission}
\label{continuum}

Figure~\ref{g_cont} shows the SMA map of the 1.3~mm continuum emission towards G24. The
inset shows the map obtained only from the very extended configuration at
1.3~mm towards G24 A1, A2, and D overlaid on the PdBI continuum map at 1.4~mm. The positions, fluxes,
and deconvolved sizes of the cores measured as the average diameter of the 50\% contour
from the SMA map are given in Table~\ref{table_cont}.

As seen in Fig.~\ref{g_cont}, all but core B have been detected at 1.3~mm. The emission
towards cores A1 and A2 has been resolved into additional cores with the higher angular
resolution provided by the very extended configuration observations (Fig.~\ref{g_cont}).
Core A1 has been resolved into 3 cores named A1, A1b, and A1c, whereas core A2 has been
resolved into two named A2 and A2b. The differences between the position and geometry of
cores A1c and A2b in the SMA VEX configuration map and the PdBI one are probably due to
the slightly extended structure of the sources and the different {\it(u,v)}-coverage of
the observations. The cores are aligned in a southeast-northwest direction coincident with
that of the molecular outflow associated with cores A1 and/or A2 (see Figs.~\ref{outflows}
and \ref{outflows_high}). The core associated with the \HC\ region is A1. To derive a
rough estimate of the free-free contribution of the embedded \HC\ region at millimeter
wavelengths, we extrapolated the optically thin free-free 7~mm continuum emission ($S_\nu
\propto \nu^{-0.1}$) measured by Beltr\'an et al.~(\cite{beltran07}). The integrated flux
density at 7~mm is 101~mJy, and the expected free-free emission at 1.3~mm is
$\sim$85~mJy.  Therefore, the dust continuum emission associated with A1 is 102~mJy (see
table~\ref{table_cont}). Note that the continuum peak intensity of core A1 is
$\sim$86~mJy/beam, which is very similar to the expected free-free emission. This means
that there is almost no dust emission within the \HC\ region.


\begin{figure*}[hbt]
\centerline{\includegraphics[angle=0,width=15cm]{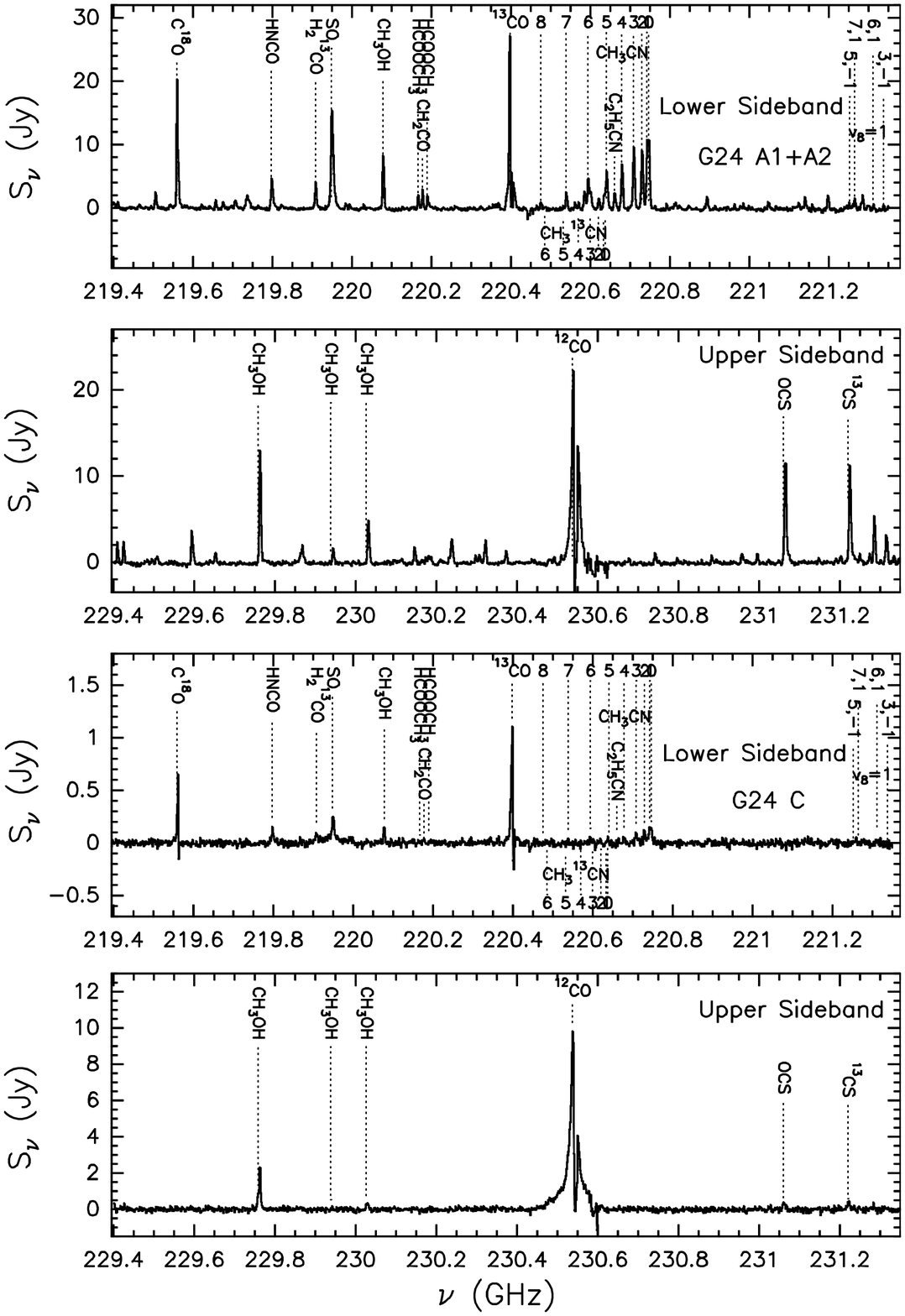}}
\caption{Upper and lower sideband SMA spectra obtained by integrating the emission 
inside the 3$\sigma$ contour level area towards G24 A1+A2 ({\it two upper
panels}) and G24~C ({\it two lower panels}). The molecular lines are
marked. The different $K$-components are marked with dashed lines in the upper
(lower) part of each spectrum in the case of \MCN\ (\MCNII). The $K,l$ quantum
numbers of the vibrationally excited ($v_8=1$) \MCN\ transitions are indicated
above the spectrum. Only the  analyzed lines are labeled.}
\label{spectra}
\end{figure*}

\begin{figure}[hbt]
\centerline{\includegraphics[angle=0,width=8.5cm]{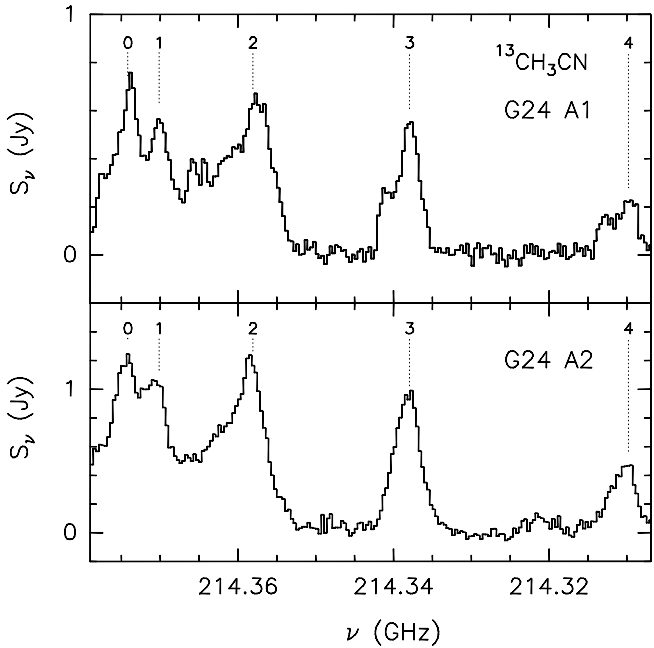}}
\caption{\MCNI~(12--11) spectrum obtained with the PdBI by integrating the emission 
inside the 3$\sigma$ contour level area towards G24 A1 and A2.
Different $K$ numbers are marked with dashed lines in the upper part of the
spectrum. The spectral resolution has been degraded to 0.5~\kms.}
\label{spectra_13ch3cn}
\end{figure}

We have also detected, although at a 3$\sigma$ noise level, core D at 1.3~mm with the SMA
and 1.4~mm with the PdBI (see Fig.~\ref{g_cont}). This core was first detected at 2.6~mm
with the PdBI and at 2~mm with the Nobeyama Millimeter Array (NMA) by Furuya et
al.~(\cite{furuya02}). Core D was not detected at 3.3 and 1.4~mm with the PdBI by
Beltr\'an et al.~(\cite{beltran05}) because the expected flux at those wavelengths fell
below the 3$\sigma$ noise level of the observations. In our case, the detection of core D
is very weak, the measured SMA flux density is 14~mJy with the VEX configuration, and 
$\sim$16~mJy with both configurations (compact plus VEX). This latter value is consistent
with the extrapolation of the fit to the spectral energy distribution by Cesaroni et
al.~(\cite{cesa03}), assuming that the source size is similar to the synthesized beam
size of the SMA observations as indicated by the continuum maps. The fact that this
source is almost completely resolved out with a resolution of $\sim$0$\farcs$84 suggests
that the source has two components: a compact core, which is what we have mapped at 1.3
and 1.4~mm, plus an extended halo whose emission is filtered out by our observations.

\begin{figure}
\centerline{\includegraphics[angle=0,width=8.5cm]{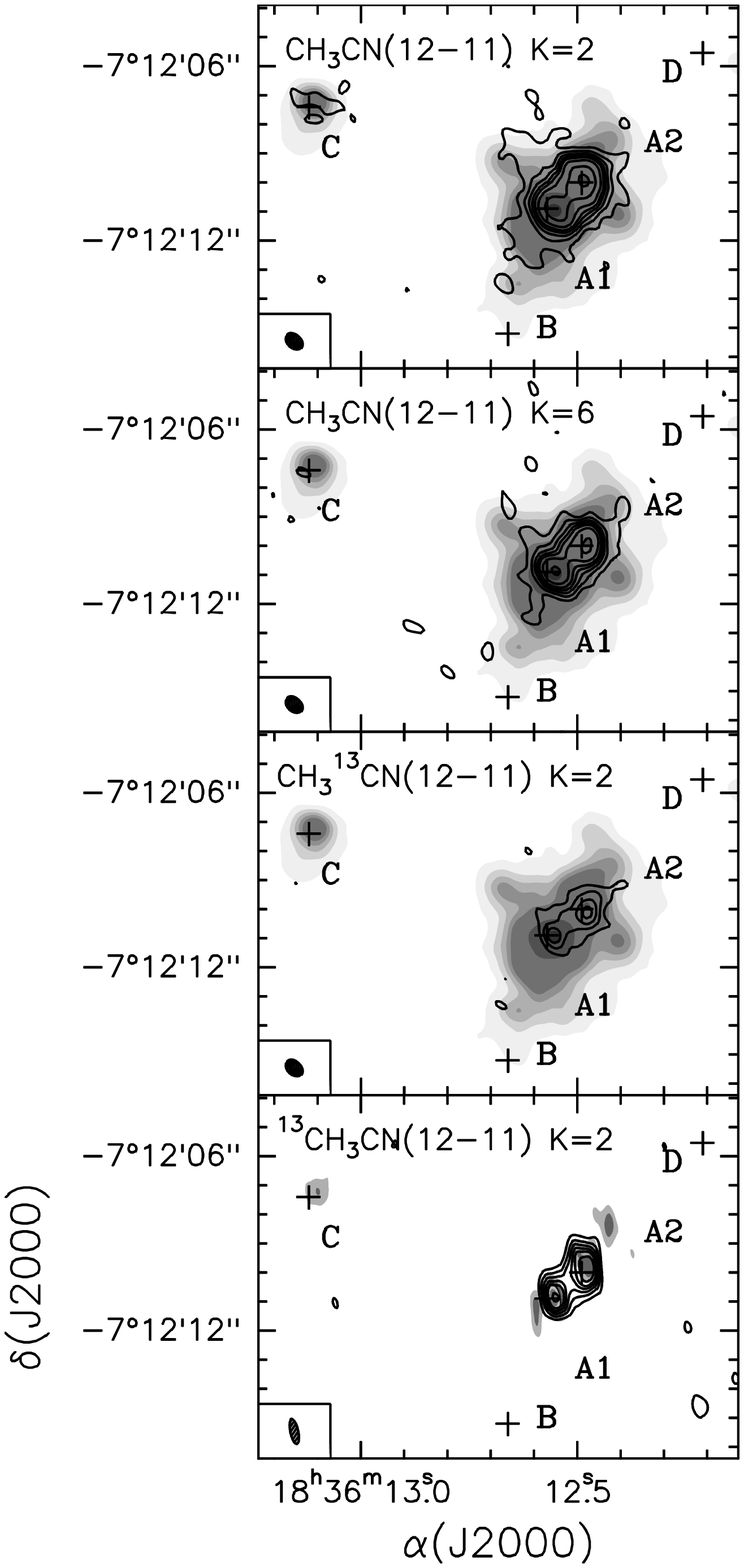}}
\caption{Overlay of the emission averaged in the velocity interval [105,
115]~\kms of the \MCN~(12--11) $K=2$, \MCN~(12--11) $K=6$, \MCNII~(12--11) K=2,
and \MCNI~(12--11) $K=2$ transitions on the maps of continuum obtained with the
SMA and PdBI ({\it bottom panel}) towards G24. Contour levels are 3, 6, 9, 12, 15, 20, 30,
and 40 times $\sigma$, where 1$\sigma$ is  15\mjy\ for all species except for
$^{13}$CH$_3$CN~(12--11) for which 1$\sigma$ is  8\mjy. The grayscale levels are the
same as in Fig.~\ref{g_cont}. The crosses mark the positions of the sources identified by
Furuya et al.~(\cite{furuya02}) and Beltr\'an et al.~(\cite{beltran04}). The
synthesized beam is shown in the lower left corner. } 
\label{dense-cont} 
\end{figure}

\subsection{\MCN\ and isotopologues}

Figure~\ref{spectra} shows the spectra of the upper and lower SMA sideband around 230 and
220~GHz, respectively, obtained by integrating the emission inside the 3$\sigma$ contour
level area towards G24 A1+A2, and G24 C. Figure~\ref{spectra_13ch3cn} shows the PdBI
spectrum around 214.3~GHz obtained by integrating the emission inside the 3$\sigma$
contour level area towards G24 A1 and A2. In Fig.~\ref{spectra}, only the lines that have
been analyzed are labeled. These are the lines with a good signal-to-noise and that could
be unambiguously identified, i.e., that are not blended with other lines. As seen in
these figures, altogether we detected 11 species (CO, HNCO, H$_2$$^{13}$CO, SO, CH$_3$OH,
HCOOCH$_3$, CH$_2$CO, CH$_3$CN, C$_2$H$_5$CN, OCS, $^{13}$CS), including 2 CO
isotopologues (C$^{18}$O and $^{13}$CO) and 2 \MCN\ isotopologues (\MCNII\ and \MCNI). 

Regarding the methyl cyanide emission, several $K$-components of \MCN~(\jdo) (up to $K$=8
for A1 and A2) are clearly detected towards G24 A1, A2, and C
(Fig.~\ref{spectra_13ch3cn}). Several $K$-components of  \MCNII\ and \MCNI~(12--11) (up
to $K$=6) are also clearly detected towards G24 A1 and A2 (Figs.~\ref{spectra} and
\ref{spectra_13ch3cn}). Figure~\ref{spectra_13ch3cn} shows only the $K$ = 0 to 4
components, which are those observed with the highest spectral resolution of 0.25~\kms.
The \MCNI~(12--11) $K$ = 5 and 6 components, which have been observed with a spectral
resolution of 3.5~\kms, have also been detected towards both cores G24 A1 and A2. Maps of
the \MCN\ (\jdo) emission averaged under the $K$ = 2 and 6 components, and of the \MCNII\
(\jdo) and \MCNI\ (\jdo) emission averaged under the $K$ = 2 component towards G24 are
shown in Fig.~\ref{dense-cont}. 

As seen in Fig.~\ref{dense-cont}, the \MCN\ and isotopologues emission is
clearly detected and resolved towards both cores A1 and A2. The emission peaks
at the same position as the millimeter continuum.

\subsection{Other high-density tracers}
 
\begin{figure*}
\centerline{\includegraphics[angle=-90,width=18cm]{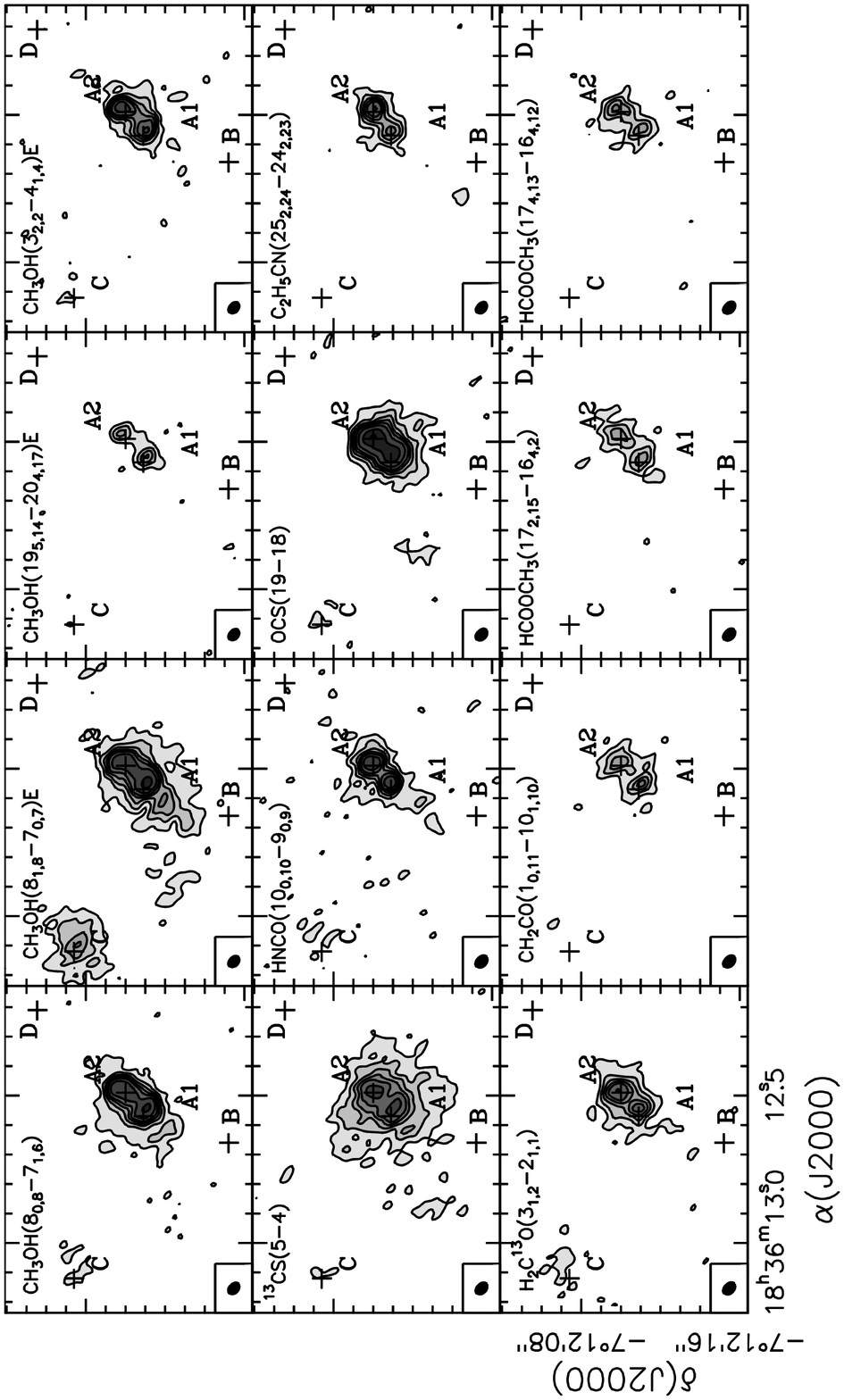}}
\caption{Images of the emission averaged in the velocity interval (105,
115)~\kms for some observed molecular species towards G24.  Contour levels are
3, 6, 9, 12, 15, 20, 30, and 40 times $\sigma$, where 1$\sigma$ is  15\mjy\ for
all species except CH$_3$OH~(8$_{1,8}$--7$_{0,7}$), $^{13}$CS~(5--4), and
OCS~(19--18) for which 1$\sigma$ is  18\mjy. The crosses mark the positions of
the sources identified by Furuya et al.~(\cite{furuya02}) and Beltr\'an et
al.~(\cite{beltran04}). The synthesized beam is shown in the lower left
corner. }
\label{dense}
\end{figure*}

In addition to \MCN\ and isotopologues, most of the species detected towards G24 are
high-density tracers. Figure~\ref{dense} shows the emission averaged in the
velocity interval (105--115)~\kms\ for the high-density tracers. The molecular
outflow tracers are discussed in the next section. As seen in this figure, all
the species have been clearly detected towards cores A1 and A2. All the
species peak towards the millimeter continuum peaks. On the
other hand, only CH$_3$OH, $^{13}$CS, HNCO, OCS, and H$_2$$^{13}$CO have been
detected towards core C, although in most cases the emission is so weak that it
is not possible to distinguish any clear emission peak. 

\begin{figure*}[hbt]
\centerline{\includegraphics[angle=270,width=18cm]{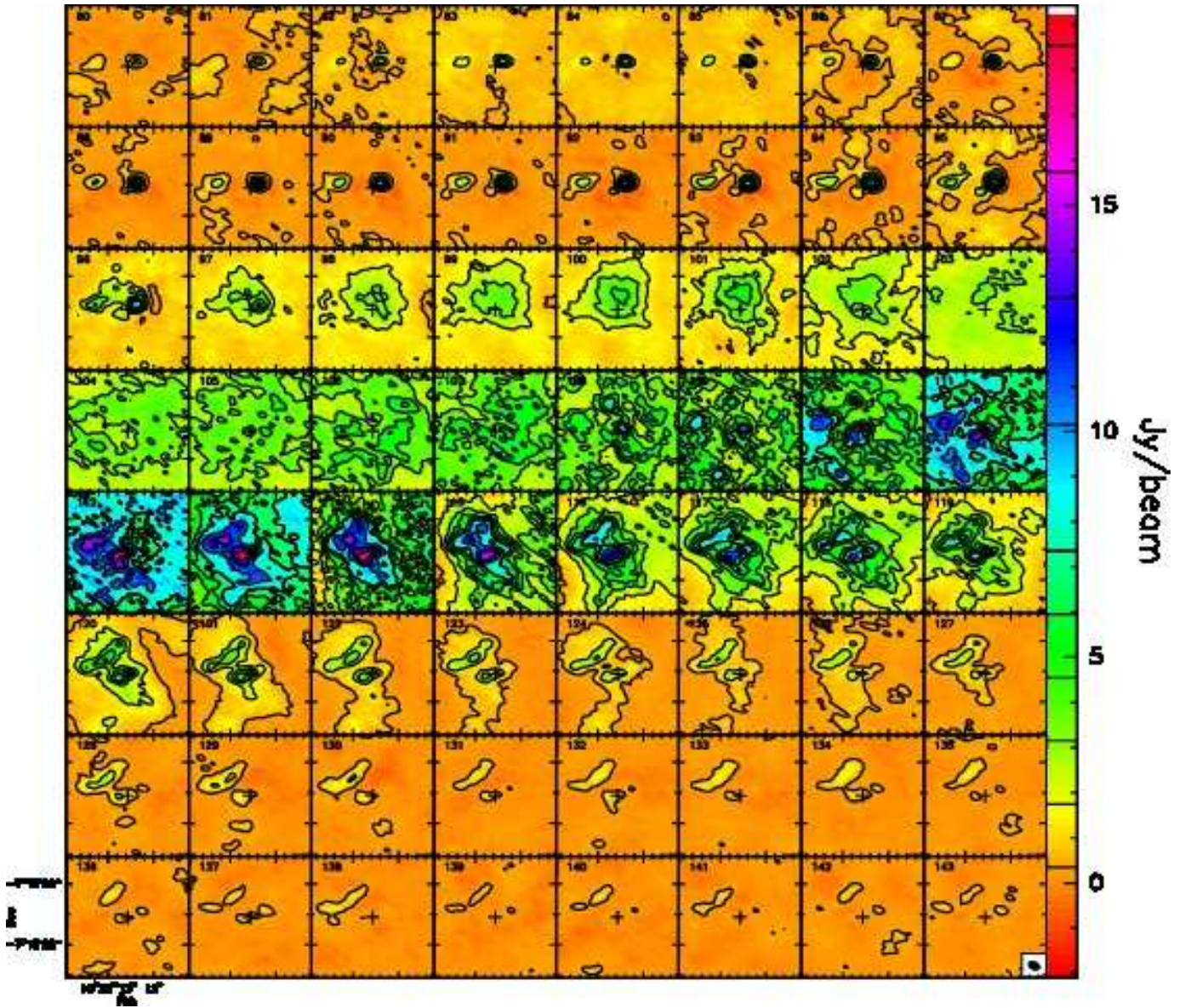}}
\caption{Combined SMA+IRAM~30-m channel maps of the $^{12}$CO~(\jdu) emission
towards G24. The contour levels range from 0.35 to 7.35~Jy~beam$^{-1}$ by steps
of 1.4~Jy~beam$^{-1}$ and from	7.35 to 18.55~Jy~beam$^{-1}$ by steps of
2.8~Jy~beam$^{-1}$. The cross indicates the phase reference center of the
SMA observations. The synthesized beam is shown in the lower right
corner of the last panel. The central velocity of each velocity interval is indicated in
the upper left corner of each panel.}
\label{chan_12}
\end{figure*}

\begin{figure*}[hbt]
\centerline{\includegraphics[angle=270,width=18cm]{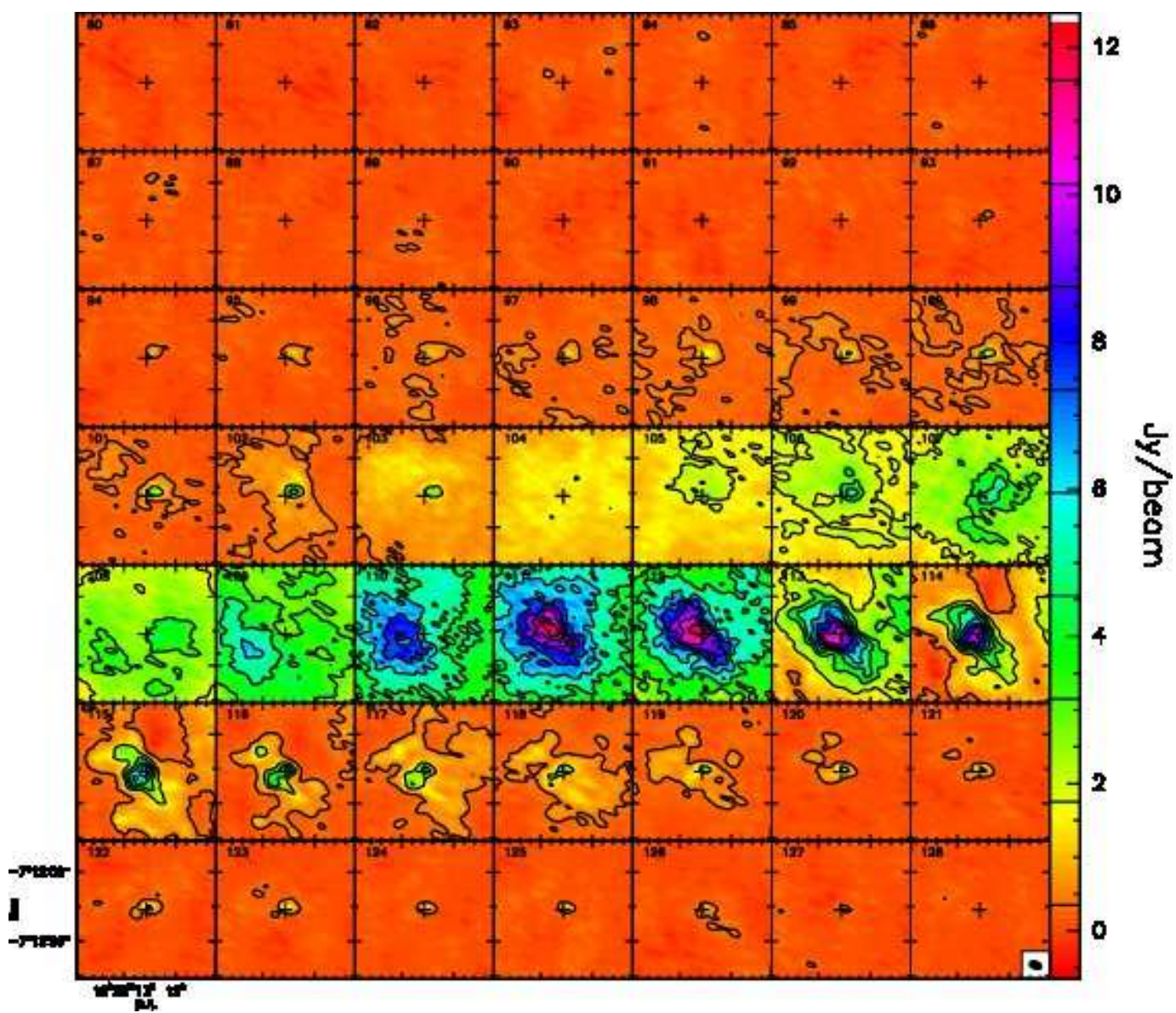}}
\caption{Same as Fig.~\ref{chan_12} for the combined SMA+IRAM~30-m channel maps of
the $^{13}$CO~(\jdu) line. The contour 
levels range from 0.35 to 11.55~Jy~beam$^{-1}$ by steps
of 1.4~Jy~beam$^{-1}$ }
\label{chan_13}
\end{figure*}

\begin{figure*}[hbt]
\centerline{\includegraphics[angle=270,width=12cm]{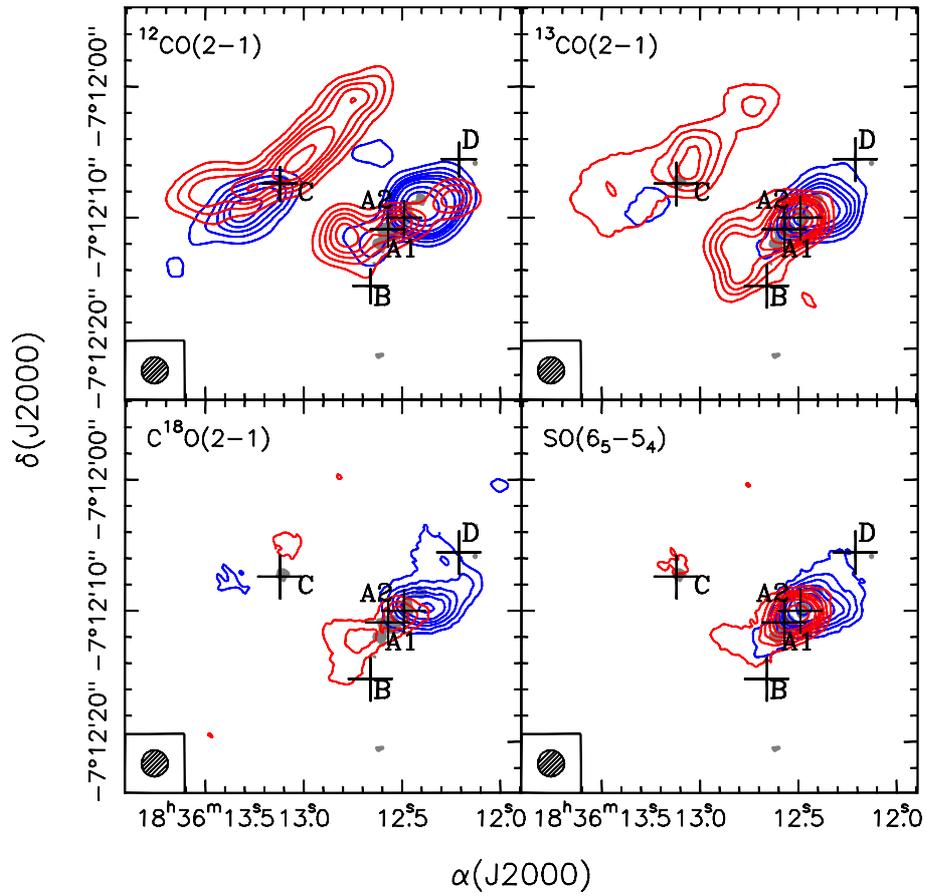}}
\caption{Blueshifted ({\it blue contours}) and redshifted ({\it
red contours}) $^{12}$CO~(\jdu),  $^{13}$CO~(\jdu), C$^{18}$O~(\jdu), and
SO~(6$_5$--5$_4$) averaged emission overlaid on the SMA VEX 1.3~mm continuum emission ({\it
grayscale}) towards G24. The maps have been reconstructed with a circular beam of
2$''$ to enhance the extended emission. The emission has been averaged in the
velocity intervals (82, 97)~\kms\ ({\it blueshifted emission}) and (127, 142)~\kms\
({\it redshifted emission}) for $^{12}$CO, (94, 103)~\kms\ ({\it blueshifted
emission}) and (116, 127)~\kms\ ({\it redshifted emission}) for $^{13}$CO, (99, 104.4)~\kms\ 
({\it blueshifted emission}) and (115.8, 122.4)~\kms\
({\it redshifted emission}) for C$^{18}$O, and
(94, 104.2)~\kms\ ({\it blueshifted emission}) and (115.6, 124.6)~\kms\
({\it redshifted emission}) for SO. Contour levels are 3, 6, 9, 12, 15, 20, 30, and 40 times
1$\sigma$, where 1$\sigma$ is 80~mJy\,beam$^{-1}$ for $^{12}$CO,  35~mJy\,beam$^{-1}$
for $^{13}$CO, 18~mJy\,beam$^{-1}$ for C$^{18}$O, and 25~mJy\,beam$^{-1}$
SO. Grayscale contour for the continuum emission is 3$\sigma$, where
1$\sigma$ is 3.2~mJy\,beam$^{-1}$. The spectral line synthesized beam is shown in the lower left
corner. The crosses mark the positions of the sources identified by Furuya et
al.~(\cite{furuya02}) and Beltr\'an et al.~(\cite{beltran04}).} 
\label{outflows}
\end{figure*}

\begin{figure}[hbt]
\centerline{\includegraphics[angle=0,width=8cm]{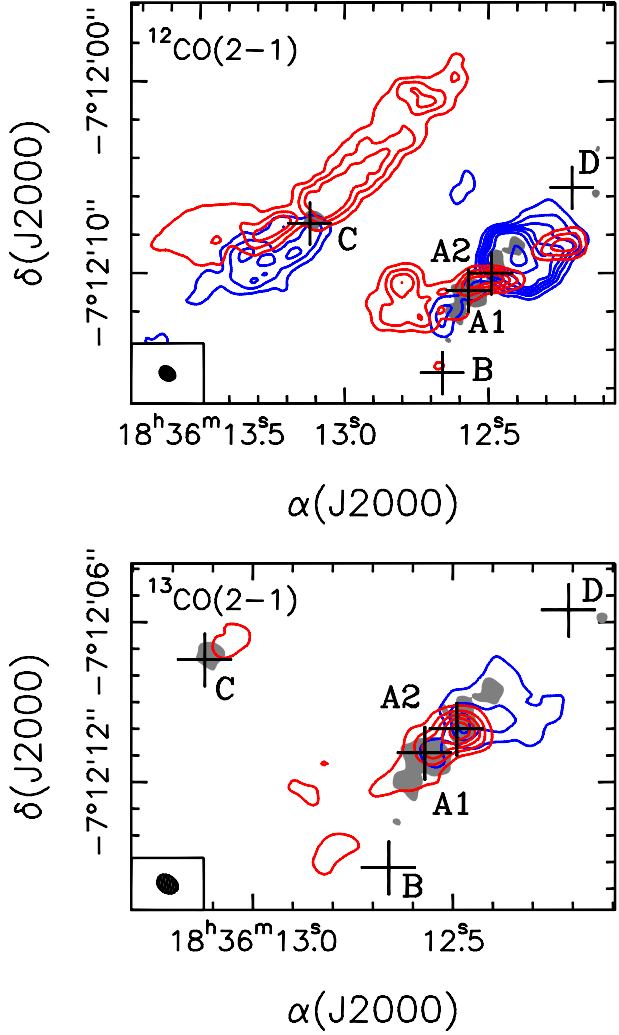}}
\caption{Blueshifted ({\it blue contours}) and redshifted ({\it
red contours}) $^{12}$CO~(\jdu) ({\it top panel}) and $^{13}$CO~(\jdu) ({\it
bottom panel}) averaged emission overlaid on the SMA VEX 1.3~mm continuum emission ({\it
grayscale}) towards G24. The synthesized beam is $0\farcs86\times0\farcs68$ at P.A.$=54\degr$
for $^{12}$CO and $0\farcs83\times0\farcs64$ at P.A.$=55\degr$ for $^{13}$CO and is shown in the lower left
corner. The emission has been averaged in the same velocity intervals as in
Fig.~\ref{outflows}. Contour levels are 3, 6, 9, 12, 15, 20, and 30 times
1$\sigma$, where 1$\sigma$ is 30~mJy\,beam$^{-1}$ for $^{12}$CO and $^{13}$CO. Grayscale contour for the continuum emission is 3$\sigma$, where
1$\sigma$ is 3.2~mJy\,beam$^{-1}$. The crosses mark the positions of the sources identified by Furuya et
al.~(\cite{furuya02}) and Beltr\'an et al.~(\cite{beltran04}).} 
\label{outflows_high}
\end{figure}


\begin{figure*}[hbt]
\centerline{\includegraphics[angle=0,width=15cm]{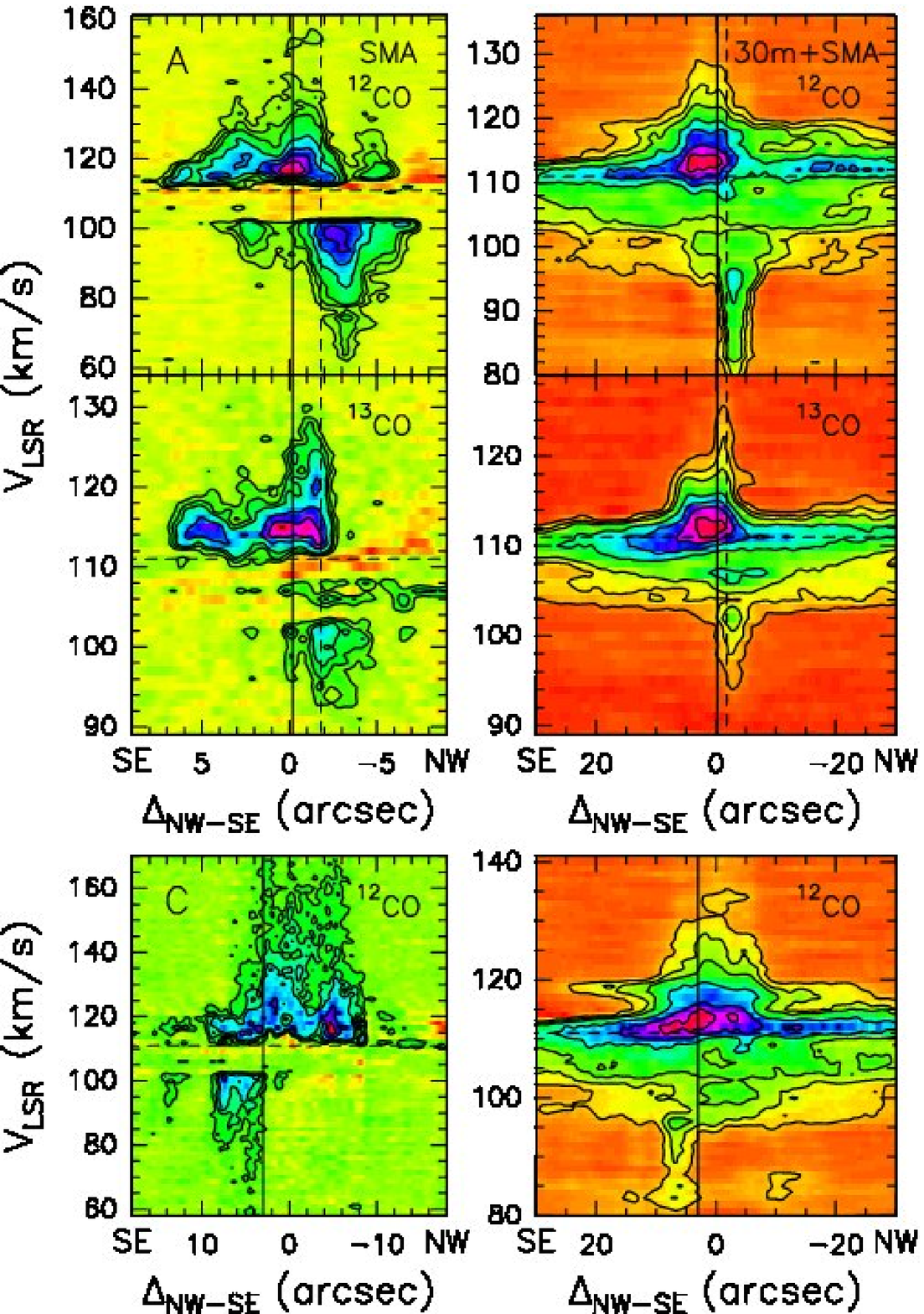}}
\caption{Position-velocity plots along the direction with P.A.=$-$40$\degr$ for outflow A
({\it top and middle panels}) and P.A.=$-$45$\degr$ for outflow C ({\it bottom panels}). The
offsets are measured from the phase center, positive towards SE. Contour levels are
3$\sigma$, 6$\sigma$ 9$\sigma$ to 45$\sigma$ by steps of 6$\sigma$ where 
$\sigma$ is 0.033~Jy\,beam$^{-1}$ for $^{12}$CO SMA, 0.33~Jy\,beam$^{-1}$ for $^{12}$CO
SMA+IRAM~30-m, 0.027~Jy\,beam$^{-1}$ for $^{13}$CO SMA, and 0.23~Jy\,beam$^{-1}$ for $^{13}$CO
SMA+IRAM~30-m. The SMA maps have the highest angular resolution. ({\it Top and middle panels}) 
The vertical solid and dashed lines indicate
the position of cores A1 and A2, respectively. The horizontal dashed line indicates the
\Vlsr. ({\it Bottom panels}) The vertical solid line indicates
the position of core C. The horizontal dashed line indicates the \Vlsr.}
\label{pv}
\end{figure*}

\subsection{Molecular outflow tracers}
\label{13co}

The $^{12}$CO~(\juz) emission towards G24 has been previously studied by Furuya et
al.~(\cite{furuya02}). These authors have discovered two bipolar molecular outflows in
the region, oriented in the same direction, northwest-southeast: one associated with
cores A1 and/or A2 and named outflow A, and the other one associated with core C and
named outflow C. The cloud systemic velocity, \Vlsr, is 111~\kms (Furuya et
al.~\cite{furuya02}). As seen in Fig.~\ref{spectra}, we have not only clearly detected
$^{12}$CO~(\jdu) towards both cores, but also the CO isotopologues $^{13}$CO~(\jdu) and
C$^{18}$O~(\jdu), and SO~(6$_5$--5$_4$). Figures~\ref{chan_12} and \ref{chan_13} show the channel maps of the $^{12}$CO and $^{13}$CO~(\jdu) SMA+IRAM~30-m
combined emission. Note that the extended emission seen in the $^{12}$CO~(\jdu)
SMA+IRAM~30-m combined channel maps for velocities between 82 and 86~\kms\ is due to a
gas component along the line-of-sight. This component is not seen in $^{13}$CO. As seen
in the channel maps, the outflow A  is clearly seen in both tracers, while the outflow C
is clearly visible in $^{12}$CO and only marginally in $^{13}$CO.

Figure~\ref{outflows} shows the $^{12}$CO, $^{13}$CO, C$^{18}$O~(\jdu), and SO
high-velocity blueshifted and redshifted averaged emission observed with the
SMA. The maps shown in this figure have been reconstructed with a circular beam
of 2$''$ to enhance the extended emission. As seen in this figure, the outflow
A, which has a P.A.=140$\degr$, is clearly seen in all lines. On the other hand,
the outflow C, which has a P.A.=145$\degr$, is clearly visible in $^{12}$CO and
$^{13}$CO and just barely visible in C$^{18}$O~(\jdu). The high-velocity
SO~(6$_5$--5$_4$) emission is hardly seen  at redshifted velocities towards core
C. 

As seen in Fig.~\ref{outflows}, the molecular outflow A shows a very prominent and poorly
collimated blueshifted lobe and a more collimated redshifted lobe. This is especially
visible in the higher angular resolution $^{12}$CO map visible in
Fig.~\ref{outflows_high}. 
The redshifted lobe shows an extension towards the northwest and the 
blueshifted one towards the southeast, especially in $^{12}$CO. The NW redshifted
emission, detected at velocities close to the \Vlsr, is still clearly visible for
velocities higher than 40~\kms\ with respect to the \Vlsr. The redshifted NW emission can
be seen in the position-velocity plots (Fig.~\ref{pv}) from about $-6''$ to $-4''$, and
the blueshifted SE emission from about $1''$ to $3''$.   A possible explanation for this
blueshifted (redshifted) emission towards the redshifted (blueshifted) lobe of outflow A
could be that outflow A is close to the plane of the sky. Alternatively, it could
indicate the presence of a second outflow in this area. 

The high-angular resolution maps (Fig.~\ref{outflows_high}) and position-velocity plots
(Fig.~\ref{pv}) seem to suggest that the core powering this molecular outflow is A2,
because it is located closer to the geometrical center of the outflow. However, with
these observations we cannot exclude the possibility of the existence of more than one
outflow (see Sect.~\ref{out_who} for a more detailed discussion).

The molecular outflow C is highly collimated. This can be seen in the
$^{12}$CO maps, either the 2$''$ resolution map or the high-angular resolution one.
As clearly seen in Figs.~\ref{outflows}, \ref{outflows_high}, and \ref{pv_episodic}, the blueshifted
lobe is much smaller than the redshifted one. This is particularly evident in the
blueshifted emission maps averaged for high outflow velocities (not shown here),
which we considered as those higher or lower than 36~\kms\ with respect to the \Vlsr. For this high
outflow-velocity range, the size of the blueshifted lobe is $\sim$3 times smaller
than that of the redshifted lobe. Taking into account the fact that the molecular
outflow consists of ambient gas that is swept up from a position close to where it
is observed, a possible explanation for the morphology of the outflow C could be
that the blueshifted lobe is close to the edge of the molecular cloud.  That is,
the blueshifted wind passes into a region where there is no, or little, molecular
material to be swept up.

Another feature of the outflow C is the redshifted emission seen southeastern of
core C. This redshifted emission is  like a finger pointing towards the east,
very prominent for outflow velocities close to the \Vlsr, but still visible for
outflow velocities higher than 40~\kms\ with respect to the \Vlsr. This feature
has a length of $\sim$8$\farcs$4 ($\sim$0.31~pc). A possible explanation for
this SE redshifted feature could be that the molecular outflow southwards of
core C is interacting with the ambient cloud close to the position of the core.
As a result of this interaction, the molecular outflow would be deflected
towards the east and the emission would shift from blueshifted to redshifted. A
similar scenario has been explained by Beltr\'an et al.~(\cite{beltran02}) 
for the intermediate-mass outflow powered by
IRAS~21391+5802 in
terms of the shocked cloudlet model scenario, where an inverted bow shock is
produced in the windward side of a dense ambient clump
(Schwartz~\cite{schwartz78}). This scenario could also explain the smaller size of the
blueshifted lobe with respect to the redshifted one, because part of the
material that would be entrained by the molecular outflow southwards of core C,
would be deflected and redshifted. We have searched for a possible high-density
southern clump responsible of the deflection of the outflow and have found a
clump in CH$_3$OH (8$_{1,8}$--7$_{0,7}$)E at a position $\alpha$(J2000)=18$^{\rm
h}$ 36$^{\rm m}$ 13$\farcs$09, $\delta$(J2000)=$-$07$^{\rm h}$ 12$^{\rm m}$
7$\fs$7. However, the clump is only seen for redshifted velocities,
$\sim$2--3~\kms\ with respect to the \Vlsr, while one would expect the clump to
be blueshifted like the southern lobe before deflection.  
Another possible  explanation for the SE redshifted emission could be the
presence of a second outflow, powered by a secondary source located close to core C but
not resolved by our observations, but the blueshifted lobe would be missing. 

Taking the blueshifted and redshifted emission into account, the total extent of
the outflows is $\sim$$13\farcs5$ ($\sim$0.5~pc) for outflow A and
$\sim$$15\farcs2$ ($\sim$0.57~pc) for outflow C.

\begin{figure}[hbt]
\centerline{\includegraphics[angle=0,width=8.8cm]{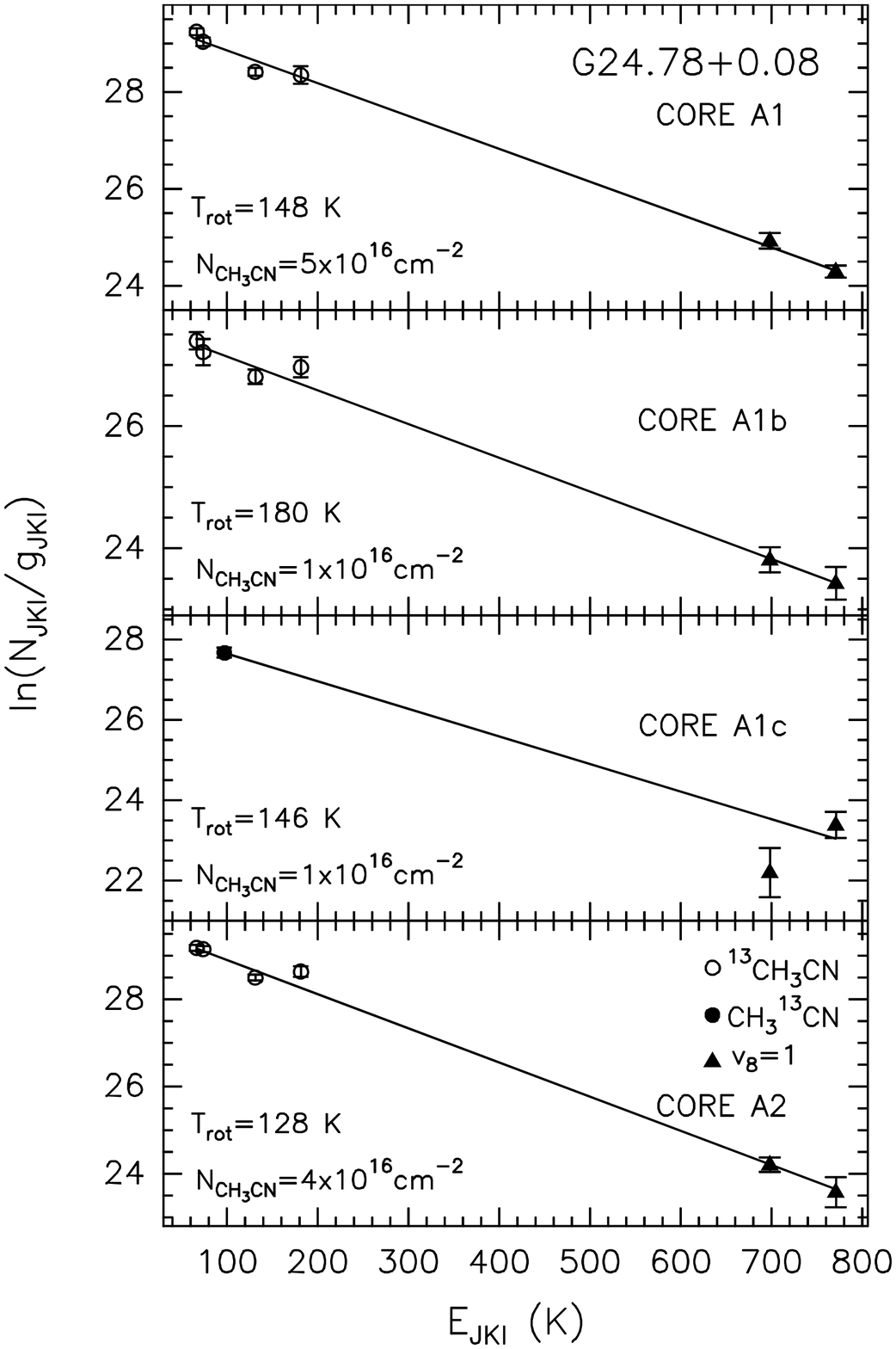}}
\caption{Rotation diagram for cores A1, A1b, A1c, and A2 in G24 with superimposed
fit. Open circles and filled triangles correspond to the \MCNI~(\jdo) and the \MCN~(\jdo) $v_8=1$ 
transitions, respectively.}
\label{boltz}
\end{figure}

\begin{figure}[hbt]
\centerline{\includegraphics[angle=-90,width=8.8cm]{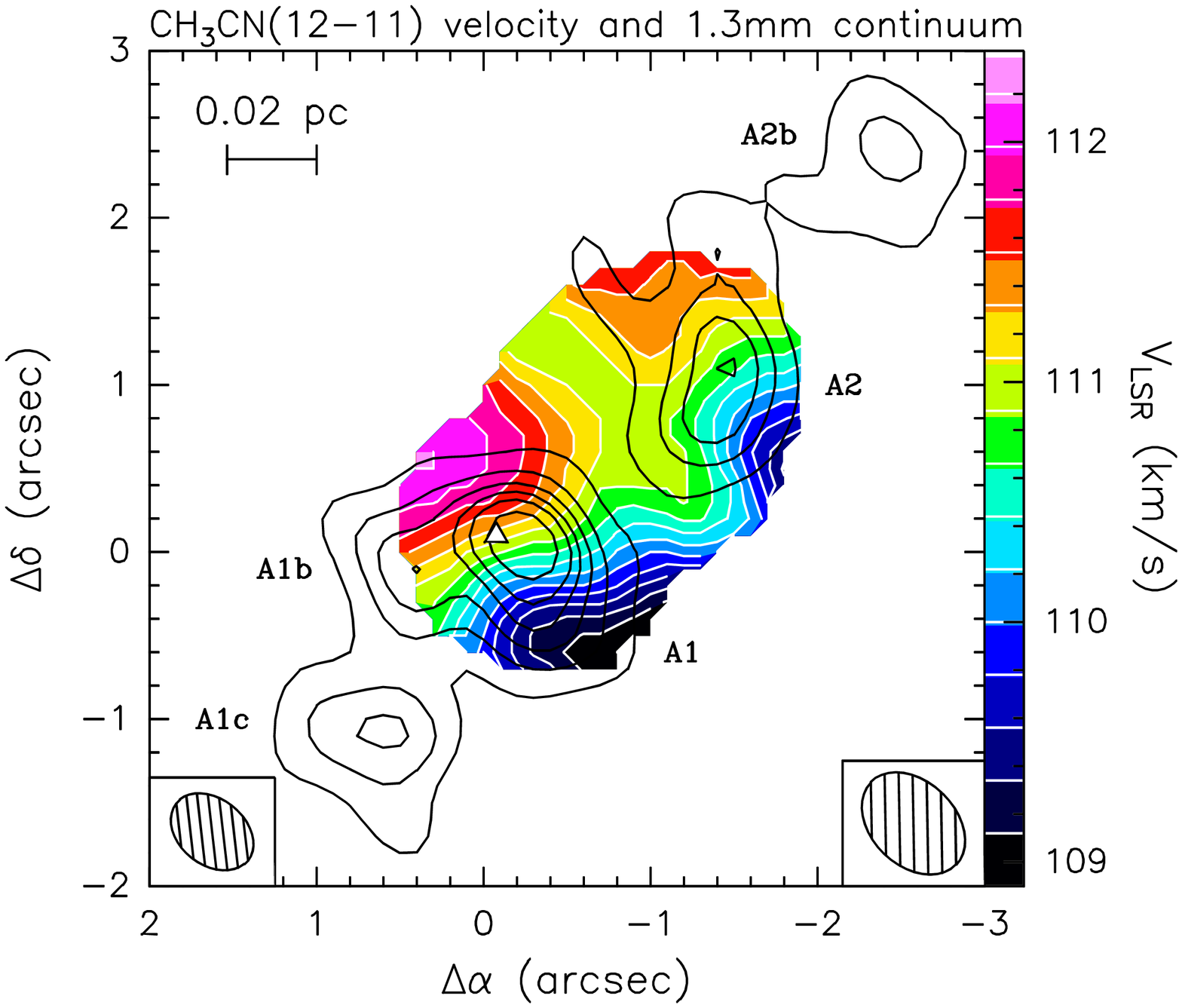}}
\caption{Overlay of the map of the 1.3~mm continuum emission obtained with
the very extended SMA configuration ({\it contours}) on that of the \MCN~(\jdo)
line velocity ({\it colour scale}). Contour levels are the same as in
Fig.~\ref{g_cont}. Offsets are measured with respect to the phase center. The triangle marks the position of the \HC\ region in G24 A1 (Beltr\'an et
al.~\cite{beltran07}). The synthesized beam of the continuum (line) emission is 
shown in the lower left (right) corner.}
\label{velgrad}
\end{figure}

\section{Analysis and Discussion}
\label{discuss}

\subsection{Temperature and mass estimates}
\label{rot}

CH$_3$CN is a symmetric top molecule, so it can be used to estimate the
temperature in the cores. The rotational temperature, $T_{\rm rot}$, and the
total methyl cyanide column density, $N_{\rm CH_3CN}$, can be estimated by means
of the rotation diagram method, which assumes that the molecular levels are
populated according to LTE conditions at a single temperature $T_{\rm rot}$. In
the high-density limit where level populations are thermalized, one expects that
$T_{\rm rot} = T_{\rm kin}$, the kinetic temperature. The CH$_3$CN ground-level
transitions appear to be optically thick for all cores, as suggested by the
ratio between the main species and isotopologues. Therefore, the Boltzmann plot 
was only performed using the $^{13}$CH$_3$CN and vibrational excited CH$_3$CN
transitions (see Fig.~\ref{boltz}). The core A1c has not been detected in
$^{13}$CH$_3$CN and only marginally in \MCNII. Therefore, the rotation diagram 
in Fig.~\ref{boltz} has been calculated with only three measurements, one
corresponding to \MCNII\ $K$=2, and two corresponding to \MCN\ $v_8=1$,
($K,l$)=(3,$-1$) and  ($K,l$)=(6,$+1$). The $T_{\rm rot}$ estimated  for this core is
146~K, and $N_{\rm CH_3CN}$ is about 1$\times$10$^{16}$~cm$^{-2}$. The relative
abundance [CH$_3$CN]/[$^{13}$CH$_3$CN] was estimated to be 46
following Wilson \&
Rood~(\cite{wilson94}) for a galactocentric distance of 3.7~kpc. The temperatures and column densities
obtained are consistent with those estimated by Beltr\'an et
al.~(\cite{beltran05}) from \MCNII\ (\jdo) and (\jcc), and \MCN\ (\jcc) $v_8=1$.

The masses of the individual cores, which are given in Table~\ref{table_cont}, have been
estimated from the 1.3~mm dust continuum emission observed with the SMA at the highest
angular resolution (VEX) assuming a dust opacity of $\simeq0.8$~cm$^{2}$\,g$^{-1}$ at
1.3~mm (Ossenkopf \& Henning~\cite{ossen94}), a gas-to-dust ratio of 100, and a dust
temperature equal to $T_{\rm rot}$. For the three cores for which an estimate of $T_{\rm
rot}$ is not possible with the rotation diagram, a range of masses is given assuming that
the dust temperature of the cores ranges from 30 to 60~K. The lower value is the kinetic
temperature of $\sim$30~K estimated from ammonia observations carried out with the Very
Large Array (VLA) by Codella et al.~(\cite{codella97}) towards G24. However, taking into
account that \MCN\ has been detected towards core C (Fig.~\ref{dense-cont}), we expect
that the temperature, at least for this core, is higher. In Table~\ref{table_cont} we
also give the masses estimated from the compact plus VEX configuration SMA map. In this
case, the emission of some individual cores cannot be resolved, and therefore, the masses
correspond to that of cores A1+A1b+A1c and A2+A2b. 


\subsection{Dense gas kinematics}
\label{kine}

Following Beltr\'an et al.~(\cite{beltran04}, \cite{beltran05},
\cite{beltran11}), the velocity field of the cores has been studied thanks to
the \MCN\ and isotopologues emission.

\subsubsection{Velocity gradients}

Figure~\ref{velgrad} shows the map of the \MCN\ line velocity for the G24 A1 and
A2 cores overlaid on the VEX continuum emission. The line velocity map has been
obtained by simultaneously fitting the \MCN\ $K=0$,1,2,3, and 4 components,
assuming identical line widths and fixing their separations to the laboratory
values, at each position where \MCN\ is detected. The validity of the assumption
of identical line widths has been checked by fitting separately different
$K$-components. Beltr\'an et al.~(\cite{beltran05}) calculated very small
variations in the \MCN~(\jdo) line widths from $K=0$ to 6 (see their
Table~7). As seen in Fig.~\ref{velgrad}, we have clearly detected the same
velocity gradients in the southwest-northeast direction observed by Beltr\'an et
al.~(\cite{beltran04}) towards G24 A1 and A2.  

The continuum emission of core A1 peaks at the center of the southern velocity
gradient, which is very close to the position of the \HC\ region. The P.A.\ of this
velocity gradient is about 50$\degr$. The velocity shift measured over an extent of
$\sim$13000~AU is $\sim$3.4~\kms, that is, a velocity gradient of about
54~\kms~pc$^{-1}$. The peak of the continuum emission of core A2  is close to the
center  of the northern velocity gradient, which has a P.A.\ of $\sim$40$\degr$.
The velocity shift measured over an extent of $\sim$11500~AU is $\sim$2.4~\kms,
that is, a velocity gradient of about 43~\kms~pc$^{-1}$.

Unfortunately our data do not have enough sensitivity to allow us to properly map
the velocity field towards the core C and better constrain the direction of the
velocity gradient detected in CS~(3--2) by Beltr\'an et al.~(\cite{beltran04}).

\subsubsection{Multiple components towards G24 A1}

\begin{figure}[hbt]
\centerline{\includegraphics[angle=0,width=8.7cm]{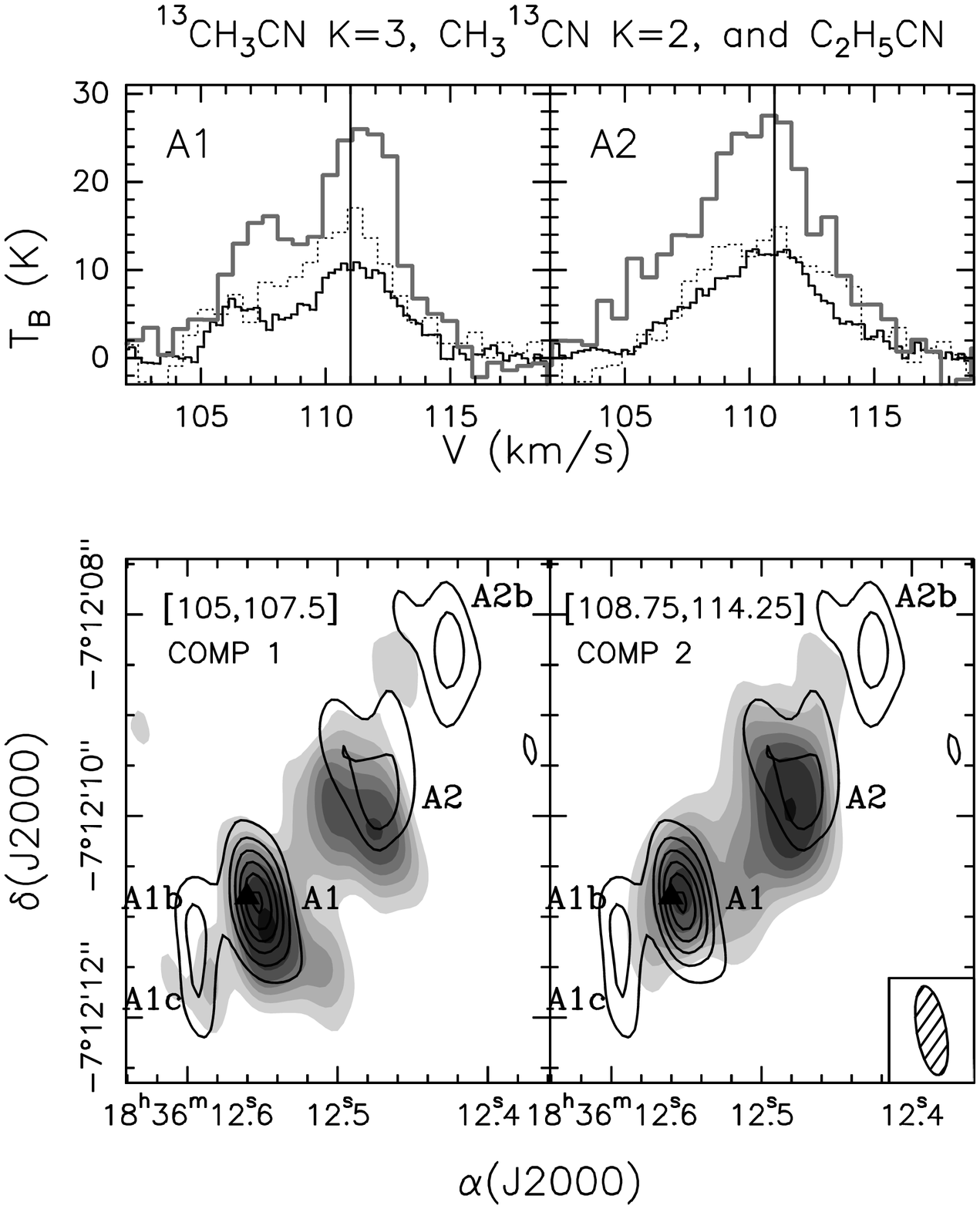}}
\caption{({\it Top panels}) \MCNI\ (\jdo) $K$=3 ({\it black line}), \MCNII\ (\jdo)
$K$=2 ({\it dotted line}), and C$_2$H$_5$CN~(25$_{2,24}$--24$_{2,23}$) ({\it gray line})
 obtained by integrating the emission inside the 3$\sigma$  contour level area 
towards G24 A1 ({\it left}) and  A2 ({\it right}). The vertical solid line indicates
the \Vlsr. ({\it Bottom panels}) Overlay of the averaged emission in the velocity interval (105,
107.5)~\kms (component 1; {\it left})  and (108.75, 114.25)~\kms (component 2; 
{\it right}) of the \MCNI~(12--11) $K=3$ transition on the map of the 1.4~mm
continuum emission 
obtained with the PdBI towards G24 A1 and A2. Contour levels are the same as in
Fig.~\ref{g_cont}. The grayscale levels are 3, 6, 9, 12, 15, 20, and 30 times 
$\sigma$, where 1$\sigma$ is 3.7\mjy\ for component 1, and 7.2\mjy\ for component
2. The black triangle marks the position of the \HC\ region in G24 A1 (Beltr\'an et
al.~\cite{beltran07}). The synthesized beam is shown in the lower right corner.}
\label{components}
\end{figure}

The analysis of \MCNI~(\jdo) has clearly revealed the existence of 2 velocity components
towards core A1. This is clearly seen in Fig.~\ref{spectra_13ch3cn} for $K$=3 and 4. The
second velocity component is less evident for $K$=1 because it is blended with  $K$=0, and
for $K$=2 because it is blended with some species typical of hot cores, such as dimethyl
ether (CH$_3$OCH$_3$) and/or methyl formate (CH$_3$OCHO). On the other hand, the $K$=0 transition is very close to the
edge of the observed band, which makes it very difficult to distinguish the second
velocity component. The 2 velocity components can also be seen in  \MCNII\ and
C$_2$H$_5$CN as shown in Fig.~\ref{components}, where we show the spectra averaged over
the core towards A1 and A2. Note that  the reason for showing \MCNII~(\jdo) $K$=2 instead
of $K$=3 is due to the fact that the latter one is blended with \MCN~(\jdo) $K$=6. 

\begin{figure}[hbt]
\centerline{\includegraphics[angle=-90,width=8.8cm]{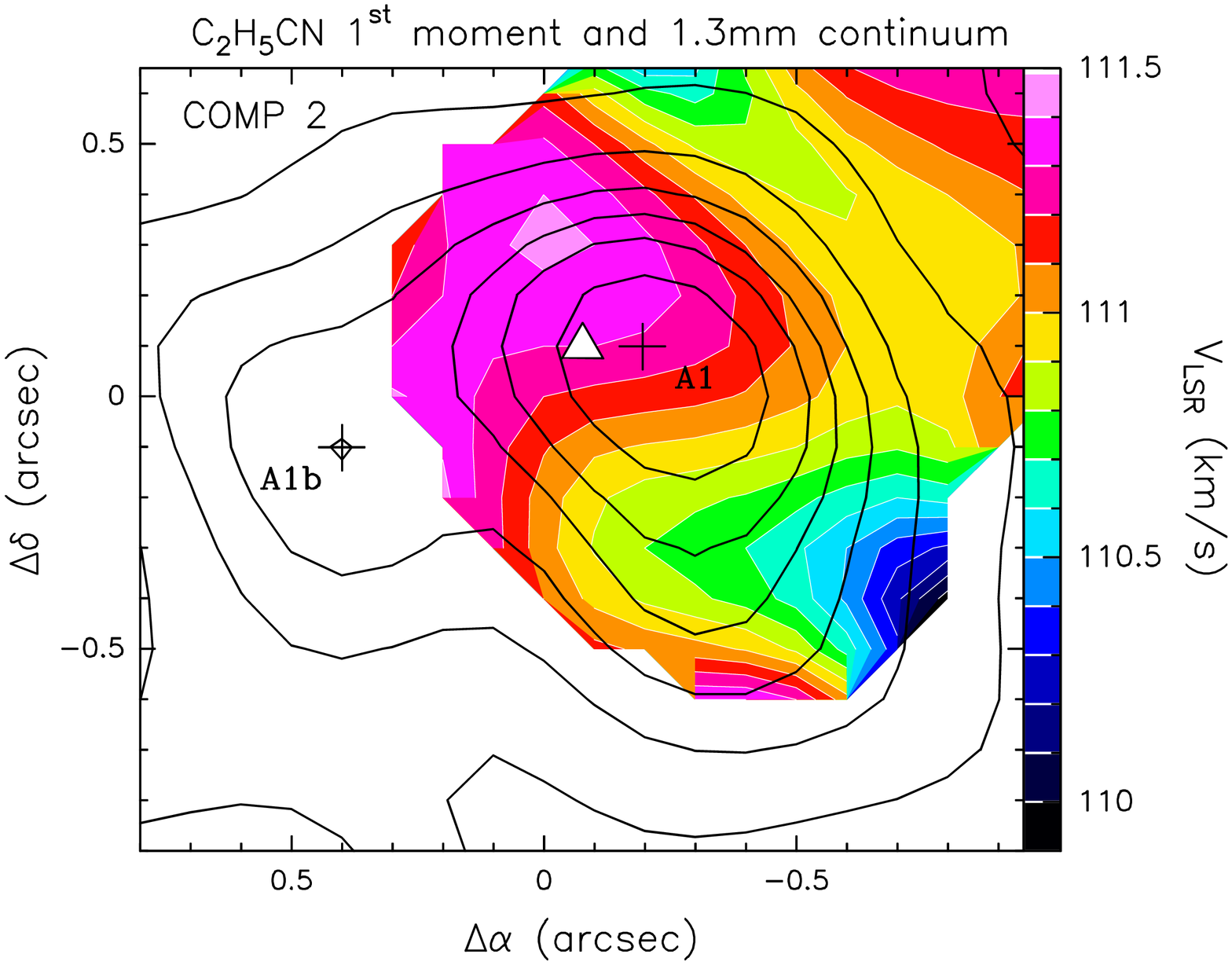}}
\caption{Overlay of the map of the 1.3~mm continuum emission obtained with
the very extended SMA configuration ({\it contours}) on the map of the first moment of
the C$_2$H$_5$CN~(25$_{2,24}$--24$_{2,23}$) second velocity component ({\it colour scale}). Contour levels are the same as in
Fig.~\ref{g_cont}. Offsets are measured with respect to the phase center. The triangle marks the position of the \HC\ region in G24 A1 (Beltr\'an et
al.~\cite{beltran07}) and the crosses the position of sources A1 and A1b. }
\label{velgrad_c2h5cn}
\end{figure}

As seen in Fig.~\ref{components}, the spectra towards A1 clearly show 2 components.
Component 1 that peaks at about 106~\kms\ (107~\kms\ for C$_2$H$_5$CN), and component 2
that peaks at about 111~\kms\ (the systemic velocity). The spectra towards A2 do not show
an apparent component peaking at a velocity of $\sim$106~\kms, although they do show a
significant blueshifted wing. The bottom panels in Fig.~\ref{components} show the \MCNI\
(\jdo) $K=3$ emission averaged for a velocity interval (105, 107.5)~\kms, corresponding to
component 1, and (108.75, 114.25)~\kms, corresponding to component 2. As seen in these
maps, the 2 components peak at a different position. Component 1 peaks towards the
southwest of the core A1 continuum emission peak, and its emission is clearly elongated in
the SW direction. On the other hand, the emission peak of component 2 coincides with that
of the core A1 continuum emission and with the position of the \HC\ region. For core A2,
the peak of the emission averaged for the 2 velocity intervals is similar although not coincident,
and in both cases is displaced from the position of the continuum emission peak. A SW
elongation is also visible for component 1, but not for component 2. Hence, we cannot
discard a priori the existence of 2 velocity components also for core A2. 

We have searched for the two different velocity components in the other high-density
tracers observed with the SMA, and in particular in \MCN\ and \MCN\ $v_8$=1. However,
although the spectral resolution of the observations is enough to disentangle the
emission of the two components, what we have found is a significant blueshifted wing in
some cases, but not two clearly separated components. We have reconstructed the PdBI
\MCNI\ $K$=3 and the SMA \MCN\ $K$=3 maps with the same synthesized beam of $0\farcs89
\times 0\farcs0.59$ P.A.=10$\degr$, i.e.\ a beam encompassing the beams of the SMA and
PdBI maps,  and have recalculated the spectrum towards A1. The component 1 was still
clearly visible in   \MCNI, even after degrading the spectral resolution to that of the
SMA observations (0.6~\kms), and in \MCNII. On the other hand, the component 1 was still
not evident in the \MCN\  spectrum.


\begin{table*}
\caption[] {Properties of the molecular outflows$^{a,b}$}
\label{tco}
\begin{tabular}{lccccccc}
\hline
&\multicolumn{1}{c}{$R^{c}\,\cos~i$}&
\multicolumn{1}{c}{$M_{\rm out}$}&
\multicolumn{1}{c}{$\dot M_{\rm out}\,\tan~i$}&
\multicolumn{1}{c}{$P^{d}\,\sin~i$}&
\multicolumn{1}{c}{$E^{d}\,\sin^2~i$}&
\multicolumn{1}{c}{$F^{d}\,\sin^2~i/\cos~i$}&
\multicolumn{1}{c}{$t_{\rm out}\,\cot~i$}
\\
\multicolumn{1}{c}{Outflow}&
\multicolumn{1}{c}{(pc)}&
\multicolumn{1}{c}{($M_\odot$)}&
\multicolumn{1}{c}{($M_\odot$ yr$^{-1}$)}&
\multicolumn{1}{c}{($M_\odot$ \kms)}&
\multicolumn{1}{c}{($10^{46}$ erg)}&
\multicolumn{1}{c}{($M_\odot \, \mbox{km s}^{-1}$\,yr$^{-1}$)}&
\multicolumn{1}{c}{(yr)} \\
\hline
A            &0.50   &27.3 &$1.5\times10^{-3}$ &245  &2.5  &$1.4\times10^{-2}$  &$1.8\times10^4$  \\
C$^{e}$            &0.57   &7.3 &$3.0\times10^{-4}$  &62  &0.6  &$2.6\times10^{-3}$  &$2.4\times10^4$  \\
C red finger$^{f}$ &0.31   &4.2 &$2.2\times10^{-4}$  &33  &0.3  &$1.7\times10^{-3}$  &$1.9\times10^4$  \\
\hline
\end{tabular}
 
(a) The tabkle indicates the estimated values. In case of knowing the inclination angle
with respect to the plane of the sky, $i$, of the outflows, the parameters should
be corrected accordingly. \\
(b) The blueshifted emission has been integrated for the velocity range (94,
103)~\kms, and the redshifted one for (116, 127)~\kms.  \\
(c) Total size of the lobes. \\		
(d) Momenta and kinetic energies are calculated relative to the cloud velocity.  \\
(e) The redshifted emission seen SE of core C has not been taken into account in the
calculations. \\
(f) Redshifted emission seen SE of core C. \\ 
\end{table*}

We have studied the velocity field of the 2 components in \MCNI\ towards A1 to see
whether the presence of another velocity component at $\sim$106~\kms\ could be 
responsible for the SW--NE velocity gradient seen in \MCN\ towards core A1 (see
Fig.~\ref{velgrad}). In fact, a velocity gradient might also be mimicked by two distinct
cores with different \Vlsr\ that are too close to be resolved by our observations. Figure~\ref{velgrad_c2h5cn} shows the first moment map of C$_2$H$_5$CN computed over a velocity
interval  from  108.5 to 114.5~\kms, that is, over the range of the second velocity
component. We have analyzed the velocity field of C$_2$H$_5$CN, because this is the
high-density tracer with 2 components clearly distinguishable that shows the
strongest emission towards G24~A1 (see Fig.~\ref{components}). The analysis indicates that
component 2 shows a velocity gradient similar to the one  observed in \MCN, centered
close to the position of the continuum emission peak and the \HC\ region, although with a
velocity range slightly smaller ($\sim$109.9--111.4~\kms). 
Therefore, we conclude that the
velocity gradient seen towards core A1, first detected in \MCN\ by Beltr\'an et
al.~(\cite{beltran04}), could only be marginally affected by the presence of a second core at a
different \Vlsr, and is produced by a structure rotating around the star powering the \HC\ region.
Hence, the velocity component at $\sim$106~\kms\ would not be related to this gradient.
The presence of a SW core is also discarded by the fact that a secondary continuum
emission peak is not detected in that direction (Fig.~\ref{components}). One would expect
that any embedded core should be better detected in continuum than in line emission.

\subsection{The molecular outflows}

Figures~\ref{outflows} and \ref{outflows_high} show the molecular outflows A and C mapped
in $^{12}$CO~(\jdu), $^{13}$CO~(\jdu), C$^{18}$O~(\jdu), and SO~(6$_5$--5$_4$) towards
G24. In Fig.~\ref{pv} we show the position-velocity plots along the outflows A and C. For
the sake of completeness we show both the SMA and the combined SMA+IRAM~30-m data. The outflow
C is hardly visible in $^{13}$CO~(\jdu) with the SMA at a resolution of $\sim$0$\farcs$7,
and hence, for this outflow we show only the $^{12}$CO~(\jdu) data. Note that the cuts
were made along P.A.=$-$40\degr\ (outflow A) and P.A.=$-$45\degr\ (outflow C) and all the
plots have been obtained after averaging the emission along the direction perpendicular
to the cut with the purpose of increasing the S/N of the plots.

\subsubsection{Which source is powering outflow A?}
\label{out_who}

One of the questions that still remains open in the study of G24 is which core, A1 or A2
(or both), is driving outflow A. The high-angular resolution maps
(Fig.~\ref{outflows_high}) seem to indicate that the core powering this outflow is A2, 
because it is located closer to the geometrical center of the outflow. This is especially
visible in the $^{13}$CO map, where outflow A shows a clear bipolarity centered in core
A2. However, the most clear evidence that A2 is the powering source comes from the
position-velocity plots along the outflow (Fig.~\ref{pv}). In these plots the vertical
solid and dashed lines indicate the position of core A1 and A2, respectively. As can be
seen in these plots, core A2 is located at the position where the emission, or the
emission at the highest outflow velocity, changes from being blueshifted to redshifted.
This is more evident in the $^{13}$CO SMA plots, because no redshifted emission is
detected at any position before that of core A2. Regarding core A1, the position-velocity
plots seem to indicate that this core is not powering any outflow. However, as shown in
Fig.~\ref{outflows_high}, the CO emission is very complex towards cores A1 and A2, with
redshifted emission towards the blueshifted lobe and the opposite. Hence, we cannot
totally discard the possibility that core A1 could be powering an additional outflow in
the region. In any case, this  possible outflow would not be the most prominent one,
which, as already mentioned, is driven by core A2.

\subsubsection{Physical parameters}
\label{out}

Table~\ref{tco} gives the parameters of outflows A and C, as well as those of
the redshifted emission feature seen SE of core C. The emission of this feature
has not been taken into account when calculating the parameters of outflow C.
The parameters have not been corrected for inclination angle, $i$, of the flow
with respect to the plane of the sky. However, we have indicated in
Table~\ref{tco} the correction for the inclination 
to be applied for each outflow parameter. The size of the lobes, mass $M_{\rm out}$, outflow
mass loss rate $\dot M_{\rm out}$, momentum $P$, kinetic energy $E$, momentum
rate in the outflows $F$, and dynamical timescale $t_{\rm out}$ were derived
from the $^{13}$CO emission for the velocity ranges indicated in
Table~\ref{tco}. The $^{13}$CO channel maps used were those reconstructed with a
2$''$ circular beam. The dynamical timescale of the blueshifted and redshifted
lobes was estimated as $t_{\rm out}=R/V_{\rm out}$, where $V_{\rm out}$ is the
difference in absolute value between the maximum blueshifted or redshifted
velocity and the systemic velocity, \Vlsr=111~\kms. The $t_{\rm out}$ of the
outflow is the maximum dynamical timescale of the two lobes.  The parameters
$R$, $M_{\rm out}$, $P$, and $E$ were calculated for the blueshifted and
redshifted lobes separately, and then added to obtain the total value. The
[$^{13}$CO]/[H$_2$] abundance ratio of $2.2\times10^{-6}$ was estimated
following Wilson and Rood~(\cite{wilson94}), and assuming an [H$_2$]/[CO]
abundance ratio of 10$^4$ (e.g.\ Scoville et al.~\cite{scoville86}). The
excitation temperature, $T_{ex}$, was estimated from the SMA+IRAM~30m $^{12}$CO
peak brightness temperature, $T_{\rm B}$, of the averaged blueshifted and
redshifted emission of the outflows separately, assuming that $T_{\rm
B}$=$J(T_{ex})$--$J(T_{\rm BG})$, where the background temperature $T_{\rm BG}$
is 2.7~K. The $T_{ex}$  values are in the range 40--50~K.

\begin{figure}[hbt]
\centerline{\includegraphics[angle=0,width=8.6cm]{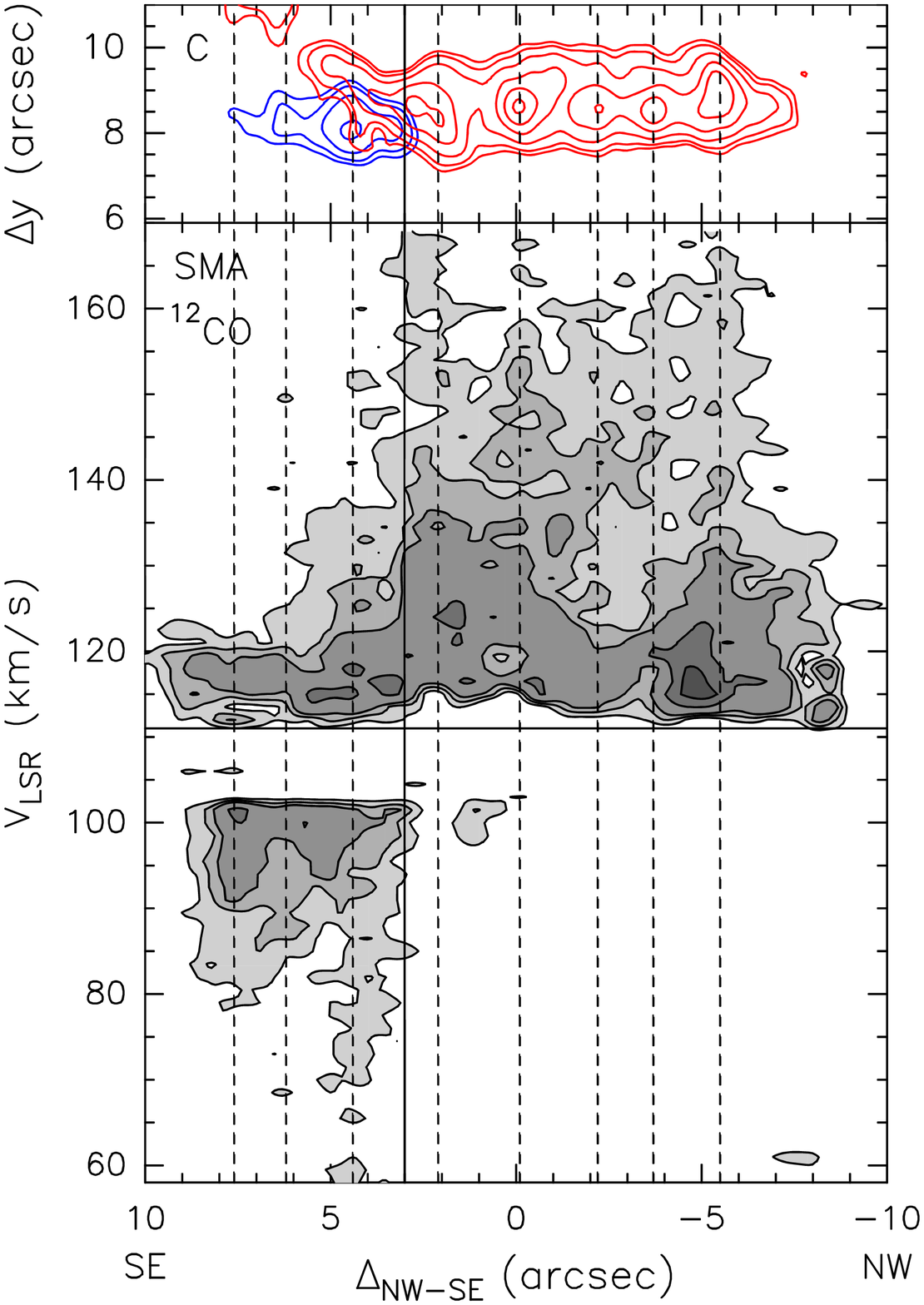}}
\caption{({\it Top panel}) Blueshifted ({\it blue contours}) and redshifted ({\it red
contours}) $^{12}$CO~(\jdu) averaged emission towards G24 C. The emission has been
averaged in  the velocity intervals (58, 74.5)~\kms\ ({\it blueshifted emission})
and (148, 167.5)~\kms\ ({\it redshifted emission}). The molecular outflow map has
been rotated $-$45$\degr$. The offsets are measured from the phase center, positive towards SE. Contour levels are 3, 4, 6, 9,
and 12 times 1$\sigma$, where 1$\sigma$ is 15~mJy\,beam$^{-1}$ (12~mJy) for
the blueshifted (redshifted) emission. ({\it Bottom panel}) Same as bottom left panel
of Fig.~\ref{pv}. The vertical and horizontal solid lines indicate
the position of core C and the \Vlsr, respectively. The position of 
each knot along the outflow axis is marked with a vertical dashed line.}
\label{pv_episodic}
\end{figure}

As seen in Table~\ref{tco}, the values of $M_{\rm out}$, $\dot M_{\rm out}$,
$P$, $E$, and $F$ of outflow A are about 4--5 times higher than those of outflow
C.  On the other hand, the size of the outflow C is larger and the dynamical
timescales are comparable. Even if we add the redshifted SE feature to the
calculations of outflow C, the values estimated are still smaller than those of
outflow A. This suggests that outflow C is less massive and powerful, probably
because it is powered by a less massive and luminous source. In fact, following
the correlations between the outflow parameters and the bolometric luminosity of
the powering source, $L_{\rm bol}$, found by L\'opez-Sepulcre et
al.~(\cite{lopez09}) for a sample of high-mass star-forming regions, the source
powering outflow A should have a $L_{\rm bol}$ of a few 10$^4$--10$^5\,
L_\odot$, while the one driving outflow C should have a $L_{\rm bol}$ of a few
10$^3$--10$^4\, L_\odot$. Furuya et al.~(\cite{furuya02}) estimated a size of
the lobes almost twice the size estimated with our observations. The fact that
the synthesized beam of their observations was $\sim$$4\farcs5$, made them more
sensitive to extended emission than ours. The $M_{\rm out}$ and $\dot M_{\rm
out}$ estimates obtained by Furuya et al.~(\cite{furuya02}) for outflow A are
three times smaller than ours, while  for outflow C their values are quite
similar to ours. The dynamical timescales are comparable.

\subsubsection{Episodic outflow}

One of the striking features about the morphology of outflow C is its clumpy
structure. This is particularly visible in Fig.~\ref{pv_episodic}, where we
show  the blueshifted and redshifted emission maps averaged for high outflow
velocities. As seen in this figure, the outflow lobes, in particular the
redshifted one, show discrete knots. The knotty appearance of the outflow  with
multiple clumps aligned along the flow axis resembles that of the high-mass
outflow HH~80--81 (Qiu \& Zhang~\cite{qiu09}), where a series of CO bullets have
been discovered. These authors propose that the bullets have not been created in
situ from entrained or swept-up ambient medium, but are the result of episodic,
disk-mediated accretion. For outflow C, the morphology could be explained in
terms of an episodic outflow where the knots are made of swept-up ambient gas.
In fact, in the position-velocity plot along the axis of this outflow, there is
an increase in the extreme emission velocity at the position of all the knots
for the lowest emission contour level. According to Arce \&
Goodman~(\cite{arce01b}), the velocity structure at the position of the knots is
characteristic of the prompt entrainment mechanism, which will produce the
highest velocities at the position of the local CO maximum, and decreasing
velocity trend towards the source. The fact that the maximum outflow velocity
does not increase linearly as a function of distance from the powering source
(Fig.~\ref{pv_episodic}) indicates that there is a different ``Hubble law"
associated with each knot, corresponding with each mass-ejection episode. The
extreme emission velocity for each mass-ejection episode will not necessarily be
the same. This is because not all the outburst from an episodic source will
necessarily have the same angle with respect to the plane of the sky or will
accelerate the ambient gas to the same maximum velocity (Arce \&
Goodman~\cite{arce01a}). 


%

\section{Conclusions}

We analyzed millimeter data obtained with the SMA and the IRAM PdBI
interferometers, and the IRAM 30-m telescope of the dust and gas emission towards the cores in the high-mass
star-forming region G24.78+0.08. The synthesized beam of $<1''$ of the
observations has allowed us to study with unprecedented high-angular resolution
the structure and the velocity field of the cores, and the molecular outflows
powered by the YSOs embedded in the HMCs.

The millimeter continuum emission towards the cores A1 and A2 has been resolved into
additional cores. Core A1 has been resolved into 3 cores named A1, A1b, and A1c, whereas
core A2 has been resolved into 2 named A2 and A2b. Interestingly, the cores are aligned
in a southeast-northwest direction coincident with that of the molecular outflows
detected in the region. This suggests a preferential direction for star formation in this
region that could possibly depend on the direction of the magnetic field. It could also
indicate star formation in filaments or sheets. The core associated with the \HC\ region
is A1. The masses of the cores for which it has been possible to estimate the rotational
temperature range from 7 to 22~$M_\odot$. The rotational temperatures range from 128 to
180~K.

The velocity gradients towards cores A1 and A2, first detected by Beltr\'an et
al.~(\cite{beltran04}) in \MCN~(\jdo), have been confirmed by the new SMA
observations. Unfortunately, the sensitivity of the observations have not allowed
us to  better constrain the direction of the velocity gradient detected in
CS~(3--2) by Beltr\'an et al.~(\cite{beltran04}).

The \MCNII, \MCNI\ and C$_2$H$_5$CN observations have revealed the existence of
2 velocity components towards A1. The component at $\sim$111~\kms\ is
associated with the velocity gradient seen in \MCN\ and peaks close to the
position of the millimeter continuum peak and of the \HC\ region. The  component
at $\sim$106~\kms\ peaks towards the SW of core A1 and does not seem to be
associated with a millimeter continuum emission peak.

The two molecular outflows in the region have been mapped in different tracers. The
position-velocity plots along outflow A and the $^{13}$CO~(\jdu) averaged blueshifted and
redshifted emission indicate that this outflow is driven by core A2. Regarding core A1,
the position-velocity plots seem to indicate that this core is not powering any outflow.
However, due to the fact that the outflow structure is very complex towards cores A1 and
A2, we cannot totally discard the possibility that core A1 could be powering an
additional outflow in the region. The outflow C is highly collimated and clumpy. The
clumpiness and the $^{12}$CO~(\jdu) position-velocity plot suggest that this outflow
could be episodic, and that the outflow clumps are made of swept-up ambient gas.

\begin{acknowledgements} 
We thank the staff of IRAM for their help during the
observations and data reduction. 
\end{acknowledgements}

\end{document}